\PassOptionsToPackage{unicode}{hyperref}
\PassOptionsToPackage{hyphens}{url}
\PassOptionsToPackage{dvipsnames,svgnames,x11names}{xcolor}
\documentclass[
  11pt,
  a4paper,
]{article}
\usepackage{xcolor}
\usepackage{amsmath,amssymb}
\usepackage{iftex}
\ifPDFTeX
  \usepackage[T1]{fontenc}
  \usepackage[utf8]{inputenc}
  \usepackage{textcomp} 
\else 
  \usepackage{fontspec} 
  \defaultfontfeatures{Scale=MatchLowercase}
  \defaultfontfeatures[\rmfamily]{Ligatures=TeX,Scale=1}
\fi
\usepackage{lmodern}
\ifPDFTeX\else
\fi
\IfFileExists{upquote.sty}{\usepackage{upquote}}{}
\IfFileExists{microtype.sty}{
  \usepackage[]{microtype}
  \UseMicrotypeSet[protrusion]{basicmath} 
}{}
\makeatletter
\@ifundefined{KOMAClassName}{
  \IfFileExists{parskip.sty}{%
    \usepackage{parskip}
  }{
    \setlength{\parindent}{0pt}
    \setlength{\parskip}{6pt plus 2pt minus 1pt}}
}{
  \KOMAoptions{parskip=half}}
\makeatother
\usepackage{longtable,booktabs,array}
\usepackage{caption}
\usepackage{calc} 
\usepackage{etoolbox}
\makeatletter
\patchcmd\longtable{\par}{\if@noskipsec\mbox{}\fi\par}{}{}
\makeatother
\IfFileExists{footnotehyper.sty}{\usepackage{footnotehyper}}{\usepackage{footnote}}
\makesavenoteenv{longtable}
\providecommand{\tightlist}{%
  \setlength{\itemsep}{0pt}\setlength{\parskip}{0pt}}

\usepackage{arxiv}
\usepackage[T1]{fontenc}
\usepackage{textcomp}
\usepackage{lmodern}
\usepackage{amsmath, amssymb}
\usepackage{graphicx}
\usepackage{booktabs}
\usepackage{longtable}
\usepackage{array}
\usepackage{etoolbox}
\usepackage{float}
\usepackage{placeins}
\usepackage{caption}
\usepackage{subcaption}
\usepackage{microtype}
\usepackage{url}
\AtBeginEnvironment{longtable}{\footnotesize}
\providecommand{\tightlist}{%
  \setlength{\itemsep}{0pt}\setlength{\parskip}{0pt}}
\usepackage{bookmark}
\IfFileExists{xurl.sty}{\usepackage{xurl}}{} 
\makeatletter
\@ifundefined{xmpquote}{}{}
\makeatother
\hypersetup{
  pdftitle={Speech-muscle coupling at jaw and ear electrodes: a volume-conductor model with orientation and electrode-count controls},
  pdfauthor={Carl Vincent Kho},
  colorlinks=true,
  linkcolor={Maroon},
  filecolor={Maroon},
  citecolor={Blue},
  urlcolor={Blue},
  pdfcreator={LaTeX via pandoc}}

\title{Speech-muscle coupling at jaw and ear electrodes: a volume-conductor model with orientation and electrode-count controls}
\author{Carl Vincent Kho \\ Independent researcher; Somach Systems, Inc., San Francisco, CA, USA \\ carl@somach.life \textperiodcentered{} https://somach.life}
\date{}

\makeatletter
\newcommand{\headlinelayoutfailure}{\PackageError{headline-layout}{Title layout drifted}{Check arxiv.sty before rebuilding}}
\patchcmd{\@toptitlebar}{\vskip 0.25in}{\vskip 0.15in}{}{\headlinelayoutfailure}
\patchcmd{\@bottomtitlebar}{\vskip 0.29in}{\vskip 0.10in}{}{\headlinelayoutfailure}
\patchcmd{\@maketitle}{\textsc{}\\}{}{}{\headlinelayoutfailure}
\patchcmd{\@maketitle}{\vskip 0.3in}{\vskip 0in}{}{\headlinelayoutfailure}
\makeatother
\AtBeginDocument{\enlargethispage{3\baselineskip}}
\begin{document}
\maketitle

\begin{abstract}

\textbf{Objective.} We compared how strongly speech-muscle sources couple to electrodes on the jaw and around the ear, and tested sensitivity to source orientation, electrode count, and anatomical detail.

\textbf{Approach.} Using reciprocity in MIDA (116 labelled compartments, 500 micrometres), we evaluated 22 electrode positions against ten segmented articulators. Muscle was modelled as both source and conducting tissue. Comparisons used equal electrode counts and a hemisphere orientation sweep, with anatomy-derived fibre models for temporalis and lateral pterygoid.

\textbf{Results.} Orbicularis oris, buccinator, mentalis, depressor anguli oris, and platysma favoured the jaw at every sampled orientation and tested electrode subset. Temporalis showed a 2.57 dB ear advantage under the uniform sweep, reduced to 1.15 dB with the derived fibre field. Its jaw-minus-ear interval [-1.45, +5.46] dB crossed zero; 50.5\% of four-site ear subsets favoured the ear. The derived lateral-pterygoid interval [-5.33, -1.12] dB favoured the ear at the fixed jaw basis. Both derived constructions have unresolved anatomical limitations. Simplifying non-muscle soft-tissue conductivities preserved montage assignments while changing gaps by up to 3.27 dB.

\textbf{Significance.} In this head model, jaw placement better captures lip and chin activity. Ear-side preferences depend on source assumptions and electrode selection. Targeted four-site placement improved lateral-pterygoid coupling by 1.07 dB over the median random ear montage. These coupling predictions require experimental testing.

\end{abstract}

\noindent\includegraphics[width=\linewidth]{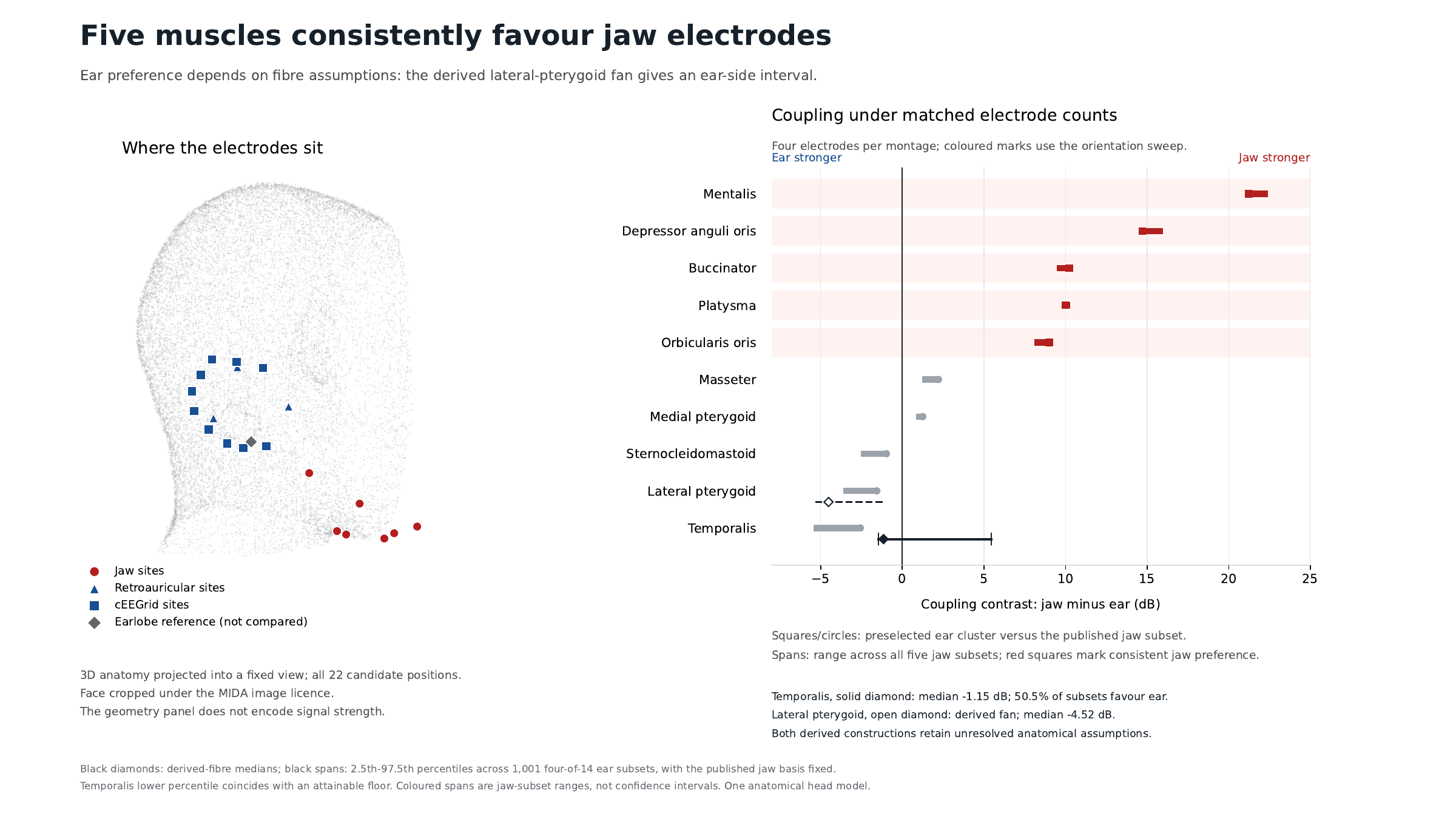}\par
\clearpage

\subsection{1. Introduction}\label{introduction}

Silent speech interfaces read articulator muscle activity from the surface of
the skin. When a word is articulated without voicing, the motor commands that
would drive the tongue, jaw, lips and hyoid still reach those muscles at
sub-threshold levels, and the resulting electrical activity is recoverable with
ordinary surface electrodes. The montage that dominates the literature places
those electrodes on the chin, under the jaw and along the throat {[}Kapur et al.
2018; Gaddy \& Klein 2020{]}, because that is where the anterior articulators are.

Ear-worn form factors have begun to appear alongside it. Retroauricular
electrode arrays demonstrably capture chewing and speaking electromyography
{[}Avramidou et al.~2024{]}, ear-mounted electrodes classify jaw clench and chew
above 90 \% accuracy {[}An et al.~2025{]}, and the cEEGrid geometry around the pinna
is an established and widely replicated wearable configuration {[}Debener et al.
2015{]}. Ear-worn devices may be easier to wear in everyday settings than
chin-mounted devices.

Forward models of muscle sources in realistic head geometry exist. HArtMuT
(Harmening et al.~2022) places roughly 3,900 muscle dipole and tripole sources
derived from MIDA's muscle segmentation, with fibre directions estimated by
principal component analysis, solved as finite-element lead fields. It is the
methodological precedent for the fibre-axis treatment used here, and its
muscle sources radiate through a homogeneous scalp compartment.

We apply this class of model to electrode placement. To test whether the
anatomically resolved conductor changes the montage comparison, we set every
non-muscle soft tissue to a single conductivity while holding geometry fixed.
Every montage assignment remained unchanged (\S{}3.5). Gap magnitudes shifted
by up to 3.27 dB, with eight of ten shifts exceeding the measurement floor.
Anatomical resolution therefore affects the reported decibel margins, even
though a homogeneous-scalp model gives the same qualitative assignments.

The design question is where to place electrodes for particular muscle
sources. The ear-EEG literature identifies the same gap in forward modelling
(Yarici et al.~2023), while ear-worn arrays already record jaw and speech
activity (Avramidou et al.~2024; An et al.~2025) using an established form
factor (Debener et al.~2015).

HArtMuT was built to answer the opposite question. Its muscle sources exist so
that muscle activity can be identified and removed from scalp EEG, and they
radiate through a homogeneous scalp compartment, a simplification its authors
state explicitly, and an appropriate one for a model whose purpose is artifact
rejection. Here, facial and cervical muscle is represented both as a source and as an
anatomically resolved conducting compartment, and the comparison concerns
sensor placement. Yarici et al.~2023 identify a related gap in the ear-EEG
literature: their review describes forward models for neural sources and ocular
artifacts but no theoretical treatment of muscle artifacts {[}Yarici
et al.~2023{]}.

We compute that coupling. Using reciprocity, the electric field is solved once
per electrode on the MIDA head model with 116 anatomically labelled
compartments, and the lead field for a muscle source at any position follows by
projection. Twenty-two electrode positions spanning the canonical jaw montage, a
retroauricular cluster and a cEEGrid C-path are compared against ten
individually segmented articulators, under isotropic and anisotropic muscle
conductivity, with an uncertainty budget that distinguishes measured terms from
unquantified limitations.

Before solving the model, we predicted that temporalis and sternocleidomastoid
would favour the ear because of their attachments at or near the temporal bone.
The prediction was recorded in the repository a day before the solve; both
commits are public and timestamped. The preliminary comparison agreed for
those two muscles and also showed an ear advantage for lateral pterygoid,
which the recorded prediction had not named. Seven of the ten articulators
favoured the jaw in that comparison. The subsequent controls did not sustain the two predicted ear advantages; the derived lateral-pterygoid interval remains on the ear side (\S{}3.1, \S{}4.2).

\begin{center}\rule{0.5\linewidth}{0.5pt}\end{center}

\subsection{2. Methods}\label{methods}

\subsubsection{2.1 Head model}\label{head-model}

The volume conductor is MIDA v1.0 (Multimodal Imaging-Based Detailed Anatomical
Model of the Human Head and Neck; Iacono et al.~2015, IT'IS Foundation, DOI
10.13099/ViP-MIDA-V1.0), segmented at 500 \textmu{}m isotropic. The ten segmented
articulator compartments and the electrode positions are shown in Figure 1.

\begin{figure}[htbp]
\centering
\includegraphics[width=\linewidth,height=0.42\textheight,keepaspectratio]{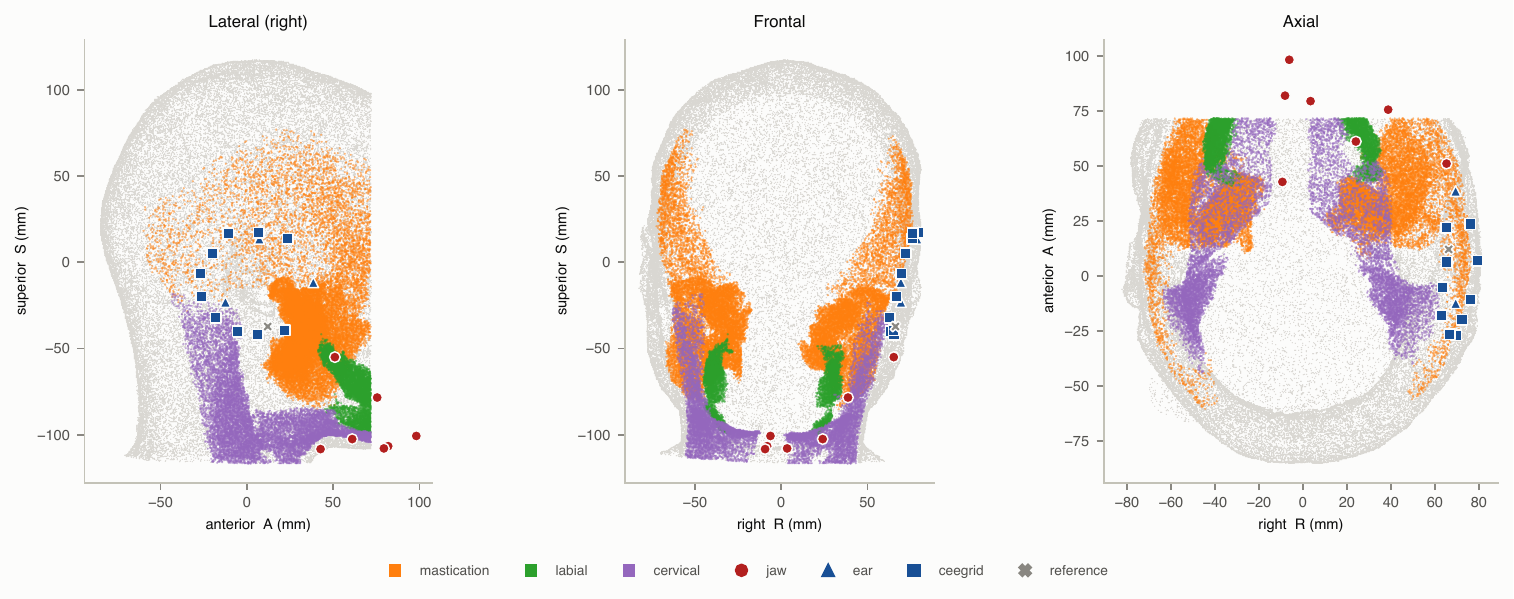}
\caption{MIDA head model with the ten segmented articulator compartments highlighted, and all 22 electrode positions shown in lateral, frontal and axial views. The face is removed anterior of the eyes at all heights, in accordance with MIDA licence clause 2.3.3.}
\end{figure}

The commonly cited count of 153 structures refers to MIDA's CAD/surface
distribution. The voxel distribution
used here contains 116 labelled structures. All results derive from that voxel
distribution; the difference in counts reflects the distributions rather than
selection of a subset.

Ten of the eighteen articulator muscles in our target set are individually
segmented and were verified present by label: masseter (66), temporalis /
temporoparietalis (63), medial pterygoid (81), lateral pterygoid (65),
orbicularis oris (75), buccinator (84), mentalis (71), depressor anguli oris
(72), platysma (60), sternocleidomastoid (68).

The suprahyoid and tongue muscles are not individually segmented. Digastric
(both bellies), stylohyoid, mylohyoid, geniohyoid, genioglossus, hyoglossus and
styloglossus are pooled into \texttt{Muscle\ (General)} (label 38; 1,975,307 voxels,
246,872 mm\textsuperscript{3}) and \texttt{Tongue} (label 42; 521,131 voxels, 65,130 mm\textsuperscript{3}). They are
therefore absent from the per-muscle results because the source segmentation
does not resolve them individually. MIDA also merges temporalis with
temporoparietalis in a single label (63) and carries the temporalis tendon
separately (98); the tendon is excluded, since its conductivity differs from
muscle and it is not a source.

\textbf{Meshing.} The labelled volume was tetrahedralised with SimNIBS 4.6's
\texttt{meshmesh} (Thielscher et al.~2015), using per-label conductivity assignment. The
finite-element electric-field solver this study repurposes for reciprocity is
described in Saturnino et al.~(2019). The base mesh contains
2,140,917 nodes and 15,415,273 elements. SimNIBS meshes the two electrodes into
it at solve time, giving 15,415,668 elements of which 12,294,182 are tetrahedra.

\textbf{Boundary.} The MIDA head model is truncated inferiorly on a planar cut, and
that face is treated as an insulating (homogeneous Neumann) boundary. The face
is fitted from 18,818 boundary triangles with unit normal {[}\textminus{}0.03336, \textminus{}0.03236,
\textminus{}0.99892{]}, tilted 2.664\textdegree{} off the RAS S axis, passing through (\textminus{}2.301, 23.112,
\textminus{}115.600) mm; residual RMS about the fit is 0.073 mm. Because the plane is
tilted it has no single S coordinate: the face spans S = \textminus{}122.07 to \textminus{}110.17 mm
across its lateral extent, and the mesh S minimum, \textminus{}122.167 mm, is a corner of
that face, not the location of the cut. All boundary clearances reported here
are perpendicular distances to this plane, not differences in S.

The plane's orientation was fitted independently of the MIDA label volume's
superior voxel axis. That axis was read from the NIfTI affine as
{[}0.03335, 0.03240, 0.99892{]} at 2.665\textdegree{} from RAS +S with a 0.500 mm step, agreeing
with the fitted normal to 0.0026\textdegree{}. The truncation therefore lies on a voxel plane
of the source segmentation. No node of the mesh lies more than 0.276 mm outside
the fitted plane, so it is the outer boundary and carries the insulating
condition.

A neck-extended variant was constructed and rejected: it does not conserve
charge, with flux failing to decay toward the domain floor (1.070 mA at S =
\textminus{}182 mm against a 1 mA injection, where the truncated control falls to 0.107 mA
at S = \textminus{}119 mm). Its terminating face is parallel to the base mesh's, displaced
70.001 mm inferiorly along the plane normal against the 70 mm extrusion
requested in \texttt{01c\_extend\_neck.py}. All published results use the truncated mesh;
the consequence for the jaw-versus-ear comparison is quantified in \S{}3.4.

\textbf{Mesh validation.} Every tag present in the mesh carries an assigned
conductivity, verified by enumeration (0 of 118 volume tags uncovered). MIDA's
own labels 100--116 lie inside SimNIBS's reserved electrode-rubber range 100--499
and are correct only because Table 1 names each explicitly; this is checked at
run time.

\subsubsection{2.2 Conductivity assignment}\label{conductivity-assignment}

Table 1 lists all conductivities and their sources. The assignments are stored
in a single file. IT'IS lookup values are evaluated at 100 Hz; the remaining assignments are
head-modelling defaults or numerical choices. We do not model conductivity
dispersion across the surface-EMG band.

Each row states its source. SimNIBS 4.6 defaults are
used for the primary head tissues, skin, fat, compact and cancellous bone, grey
and white matter, CSF, blood, eye, muscle, cartilage, air, because these are
the conventional head-modelling values and using them keeps this model
comparable with the existing EEG and tDCS forward-model literature. The IT'IS
Low Frequency database v4.2 (DOI 10.13099/VIP21000-04-2) supplies every tissue
SimNIBS carries no default for. A small number of assignments are judgement,
marked as such in Table 1, where MIDA segments a structure IT'IS does not list
separately; each carries a note giving the reasoning.

The air conductivity is chosen for numerical stability. True air conductivity is
zero, which makes the FEM system singular, so internal cavities are assigned a
small finite value. The conductivity span $\sigma$\_max/$\sigma$\_min affects numerical conditioning,
although it is not itself the stiffness matrix condition number. SimNIBS
solves iteratively. At
$\sigma$\_air = 1 \texttimes{} 10\textsuperscript{\textminus15} S/m the span reaches 1.879 \texttimes{} 10\textsuperscript{15}, the iterative solve does
not converge, and the returned fields are 10--20\texttimes{} too large while a result file
is still written. We use $\sigma$\_air = 1 \texttimes{} 10\textsuperscript{\textminus6} S/m throughout, giving a span of
1.879 \texttimes{} 10\textsuperscript{6}. A pre-flight gate refuses any assignment whose span exceeds
1 \texttimes{} 10\textsuperscript{8}.

We tested sensitivity to the air conductivity by solving at
$\sigma$\_air = 10\textsuperscript{\textminus6}, 10\textsuperscript{\textminus5}, 10\textsuperscript{\textminus4} and 10\textsuperscript{\textminus3} S/m on identical geometry. The largest
departure over the ten muscle compartments from the 10\textsuperscript{\textminus6} baseline was 0.007 dB at
10\textsuperscript{\textminus5}, 0.069 dB at 10\textsuperscript{\textminus4} and 0.467 dB at 10\textsuperscript{\textminus3}. The result is therefore
insensitive across 10\textsuperscript{\textminus6} to 10\textsuperscript{\textminus4}, where the largest departure sits roughly four
times below the measured noise floor of 0.27 dB. Departure begins at 10\textsuperscript{\textminus3}, where
it reaches 1.7\texttimes{} the floor and concentrates in the compartments nearest the oral
and nasal cavities (mentalis \textminus{}0.467, orbicularis oris \textminus{}0.433) while remote
compartments are essentially unmoved (temporalis \textminus{}0.029, lateral pterygoid
\textminus{}0.009). The operating value is the lower endpoint of the tested insensitive range,
which spans two decades.

\subsubsection{2.3 Electrode placement}\label{electrode-placement}

All 22 positions are derived from labelled anatomy and snapped to the outer
skin surface. We rejected an interactive picker because clicked coordinates
would not provide a reproducible derivation. We used placement rules so the
coordinates can be regenerated from a clean checkout and their derivation
reviewed.

Each site is defined by an anatomical anchor plus an offset in RAS millimetres,
then projected to the nearest outer-skin voxel along the surface normal.

Target localisation uses the centroid when it lies inside the target
compartment, and minimum distance otherwise. The centroid represents the
interior of a compact compartment but can fall outside a non-convex one. For
example, the mandible's centroid lies in the space enclosed by its arch; a site
placed over that point would sit over the floor of the mouth. In these cases,
we locate the site by minimum distance from the skin surface to the compartment.
The rule used for each site is recorded per row.

MIDA's head is not perfectly centred in its voxel grid. We therefore compute
the anatomical midline from the mandible label and place midline sites relative
to that plane, rather than assuming R = 0. Per-site
depth, skin surface to target compartment along the placement ray, is reported
for every site.

\texttt{pre\_tragus} is placed 14 mm anterior to the tragus, over the masseter and the
temporomandibular joint, as the retroauricular position with the shortest path
to the mastication group. The ten C-path positions are labelled \texttt{cg01}--\texttt{cg10}
after the cEEGrid layout of Debener et al.~(2015), at 12--18 mm spacing around
the pinna; all 22 sites are shown in Figure 1.

One position is withheld. \texttt{throat\_scm} is recorded as held with blank
coordinates: MIDA's sternocleidomastoid is truncated at the cut face, which
biases its centroid posteriorly, so no defensible automatic placement exists.
All downstream analyses omit this site.

The electrode model is a 10 mm diameter ellipse of 2 mm thickness, matching the
gold cup electrodes of the companion experiment.

\subsubsection{2.4 Reciprocity}\label{reciprocity}

We compute lead fields by reciprocity. For
a current dipole at position \textbf{r} with moment \textbf{p} and a recording pair (A,
B), the measured potential difference is

\[V_{AB}(\mathbf{r}, \mathbf{p}) = \mathbf{E}_{\mathrm{recip}}(\mathbf{r}) \cdot \mathbf{p} \, / \, I\]

where \textbf{E}\_recip is the field produced throughout the head by injecting current
\emph{I} between A and B. The lead field per unit injected current for a source at \textbf{r} with unit
orientation \textbf{n} is therefore E\_recip(\textbf{r}) \textperiodcentered{} \textbf{n} / I.

We use SimNIBS's tDCS solver to compute E\_recip. For each electrode, we request
1 mA between that electrode and a common reference (contralateral earlobe),
then read the resulting field inside every segmented muscle compartment. This
requires 22 solves, one per electrode, rather than the order of
10\textsuperscript{5} source-wise solves required at MIDA's resolution. The
formulations are mathematically equivalent but differ in computational cost.

Within each compartment we report the volume-weighted median \textbar E\textbar, which is
robust to the small number of very high-field elements adjacent to compartment
boundaries. For the orientation-projected montage comparisons, each site
is divided by its measured delivered-current fraction to place the fields
on a common 1 mA basis. This correction is applied during reduction and is
not repeated when the matched-count comparisons read the reduced arrays.

\subsubsection{2.5 Fibre orientation}\label{fibre-orientation}

Muscle is electrically anisotropic, but MIDA carries no fibre-direction data.
We therefore evaluate a range of orientations.

HArtMuT {[}Harmening et al.
2022{]} derived fibre directions by principal component analysis of MIDA's
segmentation to build \textasciitilde3,900 muscle sources. We follow that approach.

The isotropic condition is primary: all 22 production solves assign muscle a
single isotropic conductivity of 0.355 S/m. A second, anisotropic condition
assigns a per-element tensor with $\sigma$ = 0.4 S/m along the fibre and 0.1 S/m across
it, aligned to a PCA-derived axis.

We apply PCA only to muscles whose fibres can be represented by a single
principal axis, as in a strap-like muscle. A single axis does not describe a
sphincter (orbicularis oris), a fan (temporalis), a sheet (platysma, buccinator)
or a multi-layered muscle (masseter, lateral pterygoid). Those compartments
remain isotropic in both conditions.

Two further restrictions apply. A compartment must be individually segmented,
which excludes the six pooled suprahyoid and tongue muscles. Each axis must
also pass a bilateral mirror-symmetry test: MIDA assigns one label to both sides of a paired muscle,
so PCA on the pooled voxel cloud returns the left--right separation between the
two bellies rather than the fibre direction along either. Axes are therefore
computed per side and required to be mirror images in x. Three compartments were
tested: these were the only compartments both individually segmented and
suitable for PCA. The remaining seven were not tested.
Sternocleidomastoid
passes at \textbar dot\textbar{} = 0.98 and medial pterygoid at 1.000; mentalis fails at 0.215,
its two fragments having a right-side elongation ratio of 1.07, so no long axis
exists to find, and it receives no tensor.

The anisotropic condition therefore applies a tensor to 2 of the 10 segmented
muscles. For every other compartment, anisotropy is reported as \textbf{not applied},
because its conductivity was never varied.

The anisotropic condition is solved on the isotropic run's own mesh with only
the conductivity field replaced. SimNIBS re-meshes electrodes on each session
and two sessions of the same montage can differ, so solving on the identical
mesh removes electrode realisation, comparable in size to the effect being
measured, from the comparison.

\paragraph{2.5.1 The anatomically-constrained sweep}\label{the-anatomically-constrained-sweep}

The unconstrained sweep samples directions throughout the hemisphere. An
anatomically constrained sweep samples directions represented by the derived
fibre model. For temporalis, 8.5 \% of the derived fan reverses the
montage preference, close to the unconstrained sweep's 8.0 per cent, but only
the fan's directions incorporate the segmented anatomy. Both fractions are taken at the pre-registered cluster
{[}\texttt{results/04q\_table4\_envelope.csv}, \texttt{results/04k\_fan\_fractions.csv}{]}; the same
fan read against all fourteen ear sites gives 0.5 \%. Fractions from these two
bases describe different montage comparisons. We therefore sweep
the hemisphere first and intersect with the anatomically permitted set wherever
one can be established, reporting both fractions to show the effect of the
anatomical constraint. The main comparison adopts the derived fibre field for temporalis and the
uniform sweep for the other nine muscles. Additional fan constructions for
masseter and lateral pterygoid are disclosed in \S{}4.7, together with their
anatomical limitations.

\subsubsection{2.6 Validation}\label{validation}

We validate the solver against an analytic solution, check physical invariants
on the head mesh, and test that the validation guards execute and detect errors.

\textbf{Analytic multilayer sphere.} The full pipeline is run against a four-layer
spherical head model with a closed-form solution. Over 120 source positions
spanning radii 20--75 mm, median RDM is 4.36 \% and median MAG +4.40 \%. This layer measures absolute scale against an external analytic solution.
Multiplying every field value by a constant leaves flux radius-independence,
net boundary conservation, linearity and polarity-reversal symmetry satisfied.
A separate per-solve check compares the enclosed-current magnitude with the
requested current and rejects gross scale errors; it does not establish
absolute accuracy. Both figures are reported as validation
results. No pass criterion was fixed before these values were computed, so we report
them without assigning a retrospective threshold.

\textbf{Polarity reversal on the head mesh.} We check sign consistency on the real
geometry by solving a montage and its polarity-swapped counterpart and requiring
L(A$\rightarrow$B) = \textminus{}L(B$\rightarrow$A). On identical discretisations the residual is 7.5 \texttimes{} 10\textsuperscript{\textminus6} of
\textbar L\textbar, i.e.~6.5 \texttimes{} 10\textsuperscript{\textminus5} dB, four orders of magnitude below the measured per-site
noise floor.

\textbf{Four physical invariants}, computed from the field alone and requiring no
solver internals: (1) flux through a closed surface around an electrode is
radius-independent; (2) net current through a surface enclosing the whole domain
is zero; (3) doubling the injected current doubles the field exactly; (4)
swapping source and sink negates the field. Invariant 1 uses an exact tet-patch
integral in which flux is integrated over the interior cut of a patch of
tetrahedra using the mesh's own faces as the quadrature, so enclosure and
orientation are exact by construction (the patch surface closes to 1.2 \texttimes{} 10\textsuperscript{\textminus16}
of its own area) and no inside/outside point test is required. A stationary
plateau across radii 25--75 mm is required before any value is used. Invariants 1
and 2 run on every solve; 3 and 4 require paired solves and run on the first and
last of a batch. The guards collect all failures before raising an error, so the diagnostic
report includes every failed check.

\textbf{Convergence.} RDM against the analytic sphere was fitted across three mesh
densities, giving a convergence exponent p = 0.980, consistent with the
theoretical p $\approx$ 1 expected for a gradient quantity in a first-order FEM. This provides no goodness-of-fit evidence: three data points fitted with three
free parameters yield an exact fit by construction. The exponent p is
determined algebraically, so its agreement with theory is descriptive.

\textbf{Guard tests.} Two meta-tests protect the validation itself. A coverage test
resolves, from the abstract syntax tree, that every production script actually
invokes its guards. A synthetic-fire test requires each guard to fail on a
purpose-built input that fails that guard and nothing else, with a clean control
that trips none; this detects a guard rendered unreachable by an earlier guard's
raise, which a coverage test cannot see.

\subsubsection{2.7 Error budget}\label{error-budget}

The reported coupling comparisons are ratios between sites, montages or
muscles. A term that scales the whole lead field equally cancels in these
ratios; a term that varies between sites can affect them. Table 3 separates
these two kinds of uncertainty. Quantified terms enter the second column only
when shown to vary per site.

Table 3 also records four unresolved terms separately. Terms 1, 2, 3 and 7
have a known direction of effect on the ratio but cannot be quantified at
current precision. They carry no numeric value and do not satisfy the
admission rule for quantified per-site terms.

Electrode contact area illustrates the distinction. Each electrode's contact
is realised independently from the surface triangles beneath it, so contact
area contributes per-site variation. A uniform magnitude offset subtracts to
exactly 0 dB in any site ratio.

\subsubsection{2.8 Reproducibility and pre-registration}\label{reproducibility-and-pre-registration}

The analysis can be reproduced from a clean checkout given MIDA in
\texttt{data/} and the mesh that produced these results. Downloading MIDA
requires registration; this manual step is documented in the repository README.
MIDA cannot be redistributed under its licence and is not included.

The original mesh, \texttt{data/mida\_headneck.msh}, is not under version control, and \texttt{meshmesh}
is not deterministic: the same label volume and the same script produced
12,294,185 tetrahedra on one machine and 12,587,663 on another. A clean checkout
therefore rebuilds a mesh that is not the one these numbers came from, and the
cut-plane guard in \texttt{config.cut\_plane()} will refuse it by design, since the
plane is fitted to a specific mesh and verified by hash. Reproducing the
pipeline is possible from a clean checkout; reproducing these exact values needs
the mesh, and term 10 of the error budget quantifies what that difference is
worth.

The anatomical prediction was recorded before the model was solved. The
\texttt{expected\_at\_ear} column of the muscle configuration, predicting strong
retroauricular coupling for temporalis (``directly above ear'') and
sternocleidomastoid (``mastoid attachment''), entered the repository in commit
\texttt{fa583f6}, dated 2026-08-02. The lead-field results that test that prediction
were committed the following day, 2026-08-03. The prediction therefore precedes
the measurement by a day and by the entire solve pipeline, and both commits are
citable by hash.

Two tables in this study (\texttt{04h\_matched\_counts.csv},
\texttt{04j\_two\_axis\_verdict.csv}) were initially produced interactively and
could not be regenerated from a clean checkout. We suspected they had drifted
out of step with the renormalisation specified above, but that suspicion proved
false. An attempted correction applied the renormalisation a second time and
moved nine published numbers by up to 1.04 dB. Checking the upstream script
identified the error. This episode shows why internal consistency alone cannot
establish that a table is correct. Both tables are now generated by
\texttt{src/04h\_matched\_counts.py}, and every reported quantity is taken from
a file generated by a script in \texttt{src/}.

The original anatomical prediction did not establish a robust montage preference. Temporalis and sternocleidomastoid initially favoured the ear, as did lateral pterygoid, which the prediction had not named. The temporalis derived-field interval and the sternocleidomastoid uniform-sweep interval cross zero. The derived lateral-pterygoid interval remains on the ear side (\S{}3.1, \S{}4.7). We retain the original prediction and distinguish these conditional results rather than treating them as a uniform rejection of ear-side coupling.

\subsubsection{2.9 Ethics}\label{ethics}

This computational modelling study used a licensed anatomical model and
involved no human subjects. No IRB review was required or sought. We acquired
no new imaging, recordings or measurements from living participants.

\begin{center}\rule{0.5\linewidth}{0.5pt}\end{center}

\subsection{3. Results}\label{results}

\subsubsection{3.1 Which montage sees which muscle}\label{which-montage-sees-which-muscle}

A single gap per muscle requires a fibre direction that the model does not
provide. Comparing the best of fourteen retroauricular sites with the best
of four jaw sites also confounds placement with electrode count. We control
both comparisons. Source orientation is swept over the hemisphere at 200 directions
with the same orientation applied at both electrodes, since only a common
orientation corresponds to a physical source. Electrode count is matched by
drawing four of the fourteen ear sites at random, taking the best, and
repeating; the resulting interval says whether a preference is a property of
the montage or of which sites happen to be available.

The jaw advantage for the labial group survives both controls.
Orbicularis oris, buccinator, mentalis, depressor anguli oris and platysma
favour the jaw at all 200 sampled orientations and at every tested electrode
subsample. None of the sampled fibre directions or tested four-site selections reverses
the jaw advantage for these muscles. The full muscle-by-site
sensitivity matrix is given in Figure 2, and the per-muscle verdicts in
Figure 3.

\begin{figure}[htbp]
\centering
\includegraphics[width=\linewidth,height=0.42\textheight,keepaspectratio]{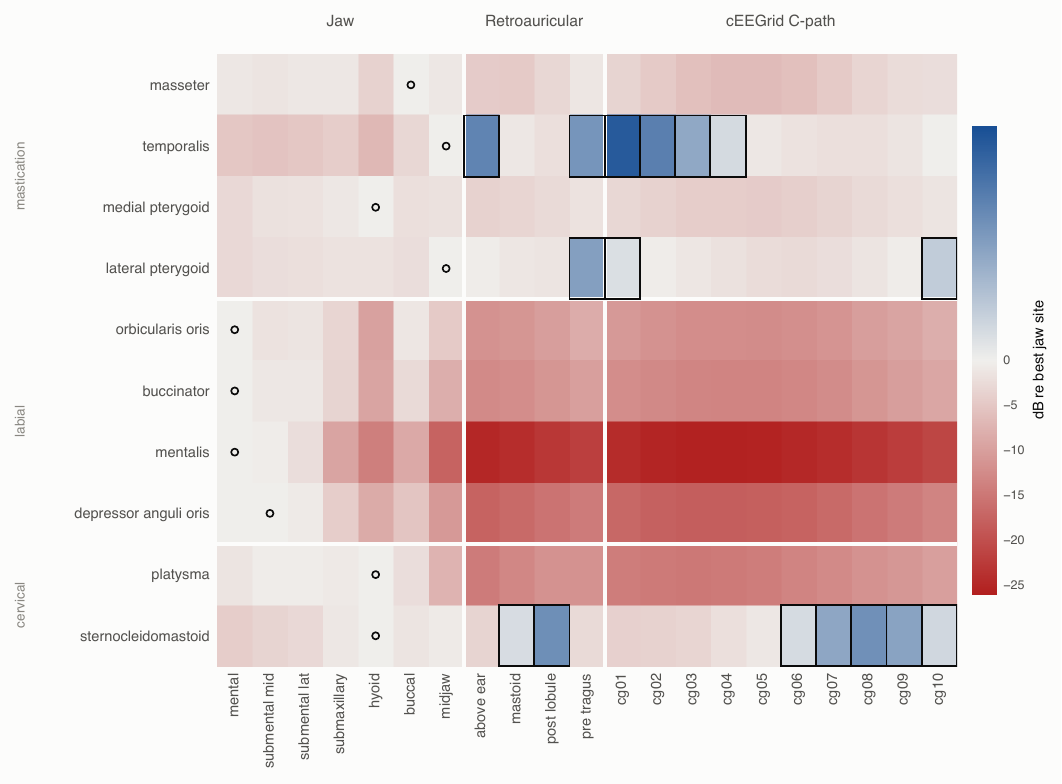}
\caption{Articulator sensitivity matrix. Median lead field in each segmented muscle compartment for each electrode site, in dB relative to that muscle's best jaw site (0 dB, ringed). Isotropic condition, truncated mesh, 22 electrodes \texttimes{} 10 muscles. The colour scale is diverging about 0 dB: red is attenuation relative to the best jaw site, blue is a site that exceeds it, and boxed cells mark every (site, muscle) pair where a retroauricular electrode beats every jaw electrode. The two arms are not equally scaled (\textminus{}28 to 0 dB against 0 to +4 dB), so colour saturation is not comparable across the midpoint; the boxes carry the sign independently of colour. The 0 dB reference here includes all jaw sites, whereas Figure 3 and Table 4 exclude those within 10 mm of the truncation face, so the named best jaw site differs for some muscles.}
\end{figure}

\begin{figure}[htbp]
\centering
\includegraphics[width=\linewidth,height=0.42\textheight,keepaspectratio]{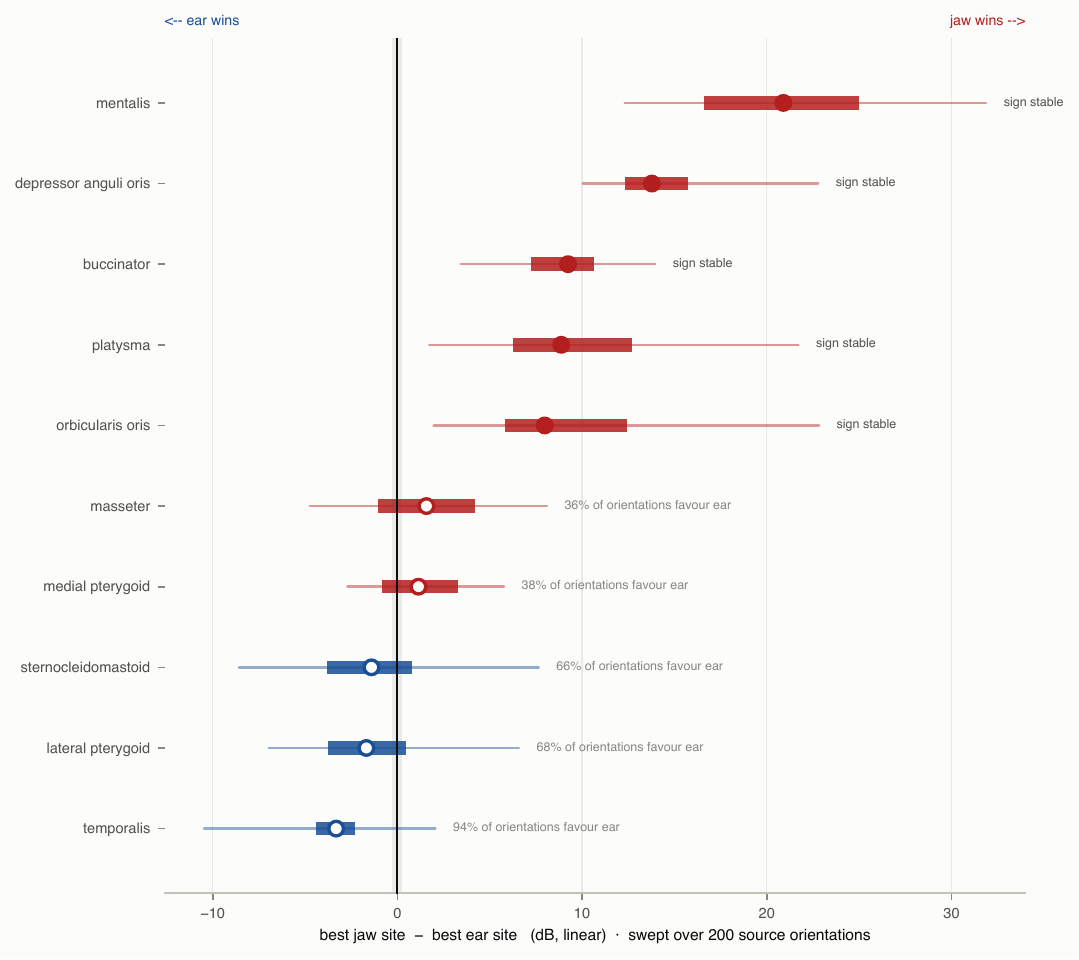}
\caption{Which montage sees which muscle. For each articulator, the difference between the best jaw site and the best retroauricular site, in dB. Negative bars are muscles the ear sees more strongly. Jaw sites within 10 mm of the truncation face are excluded. The shaded band is the measured electrode-meshing floor (0.27 dB, with the lighter band extending to its 95 \% CI upper bound of 0.65 dB). The axis is linear in dB and not a rank, because the asymmetry between the arms is itself a result: the jaw's advantages reach +20.90 dB while the ear's reach only \textminus{}3.31 dB. Both numbers are emitted from the figure's own source file by src/04m\_caption\_numbers.py, which fails the build if a caption and its figure disagree.}
\end{figure}

The temporalis preference depends on both the fibre model and site selection. Over the fibre field derived from the anatomy
(\S{}2.5.1) it reaches \textminus{}1.147 dB at the pre-registered four-site cluster, with
91.5 per cent of fibre directions agreeing, but its matched-count interval is
\textbf{{[}\textminus{}1.453, +5.458{]}} and only \textbf{50.5 per cent} of the four-site
retroauricular subsets favour the ear at all. For this derived fibre field, the sign of the temporalis contrast depends
on which four electrodes are available.

The uniform orientation sweep and the derived fibre field give different
verdicts. We use the derived field for the reported verdict. Under a
uniform orientation sweep temporalis reaches \textminus{}2.571 dB with 92.0 per cent of
orientations agreeing and an interval of {[}\textminus{}3.308, \textminus{}0.035{]} that excludes zero.
Under that treatment, temporalis would be reported as favouring the ear. The sweep assumes
source orientation is uniformly distributed over the sphere, a default
when fibre direction is unknown. For temporalis, the segmented anatomy
identifies a common insertion and permits a more specific fibre field.

For the derived temporalis field, the interval is computed by enumerating all
1001 subsets of four sites from the fourteen candidate ear sites. It therefore
has no Monte Carlo sampling error conditional on the model and these candidate
sites. The enumeration varies electrodes alone while holding the derived fibre
field fixed. The uniform-orientation intervals for the other comparisons use
random four-site draws; their Monte Carlo sampling error is not removed by this
enumeration.
Temporalis gives a median gap of \textminus{}1.147 dB with an interval of
{[}\textminus{}1.453, +5.458{]} dB, favouring the ear in 50.5 per cent of subsets
{[}\texttt{results/04p\_headline\_interval.csv}{]}.

The interval describes variation among the candidate montages rather than
uncertainty in a population estimate. We do not use it to test a null
hypothesis across the ten muscles, so multiple-comparison correction does
not apply to this descriptive comparison.

The lower bound of this interval is not a tail quantile. The most ear-favouring
outcome of any 4-site subset is the best of all 14 sites, so \textminus{}1.453 dB is a
floor fixed by the data, not by sampling, and it is attained in exactly the 28.6
per cent of subsets that contain that site, which is 4/14 as expected from
subset size alone. It coincides with the argmax gap over all 14 sites, \textminus{}1.453
dB, by construction; the two figures describe the same measurement.
The corresponding bound under the uniform orientation sweep, \textminus{}3.308 dB, is a
genuine percentile lying strictly above its own floor of \textminus{}3.314 dB, which is
attained in 2.2 per cent of draws. The two lower bounds are therefore not
comparable quantities.

The derived temporalis result uses an anatomical orientation model rather than a uniform sweep; it is not independent anatomical validation. If that derivation is rejected, the uniform-sweep result favours the ear by 2.57 dB.

The pre-registered lateral-pterygoid fan gives an interval of {[}\textminus{}5.327, \textminus{}1.115{]} dB across all 1001 four-site ear subsets at the fixed published jaw basis. Its uniform-sweep interval from random ear-subset draws is {[}\textminus{}1.59, +1.09{]} dB. These are conditional comparisons under different source models, not repeated estimates of one population parameter. The derived interval excludes zero on the ear side; it does not establish robustness across alternative fibre models or all five jaw subsets.

Rule R3 in the repository's fibre-fan pre-registration says that the derived field governs where it exists. We therefore report the derived lateral-pterygoid result alongside the derived temporalis result, with the uniform comparisons retained. The adequacy of both anatomical constructions remains a limitation (\S{}4.7).

Two other muscles have preferences that do not survive electrode subsampling.
Sternocleidomastoid (\textminus{}0.973 dB at the cluster, 60.5 per cent of orientations,
interval {[}\textminus{}1.40, +1.27{]}) and lateral pterygoid (\textminus{}1.564 dB, 65.5 per cent,
{[}\textminus{}1.59, +1.09{]}) both have intervals crossing zero. Their apparent advantages depend on which four sites are available. Reported at the unmatched argmax over fourteen sites they would read
\textminus{}1.402 and \textminus{}1.679 dB, which is why the matched comparison is the one reported.

Two favour the jaw robustly across sites but not across orientation.
Masseter and medial pterygoid have subsample intervals entirely positive, but
31.5 and 34.5 per cent of sampled orientations reverse them at the published
cluster {[}\texttt{results/04n\_site\_set\_sensitivity.csv}{]}. Table 4 reports their site robustness and orientation dependence separately.

\subsubsection{3.2 Anisotropy changes the field but not the comparison}\label{anisotropy-changes-the-field-but-not-the-comparison}

The measured effects of anisotropy on the reported jaw-versus-ear ratios
are below the electrode-meshing floor (Figure 4).
Applying a fibre tensor changes the jaw-versus-ear gap by \textminus{}0.085 dB for
sternocleidomastoid, \textminus{}0.010 dB for medial pterygoid, +0.137 dB for temporalis
and +0.036 dB for lateral pterygoid. Each change lies below the 0.27 dB
measured electrode-meshing floor, including for the two compartments that
carry a tensor. Anisotropy raises the absolute lead field substantially, by
roughly 5 dB in medial pterygoid, but it does so at the jaw and ear sites
alike, so the reported site-to-site ratios change little.

\begin{figure}[htbp]
\centering
\includegraphics[width=\linewidth,height=0.42\textheight,keepaspectratio]{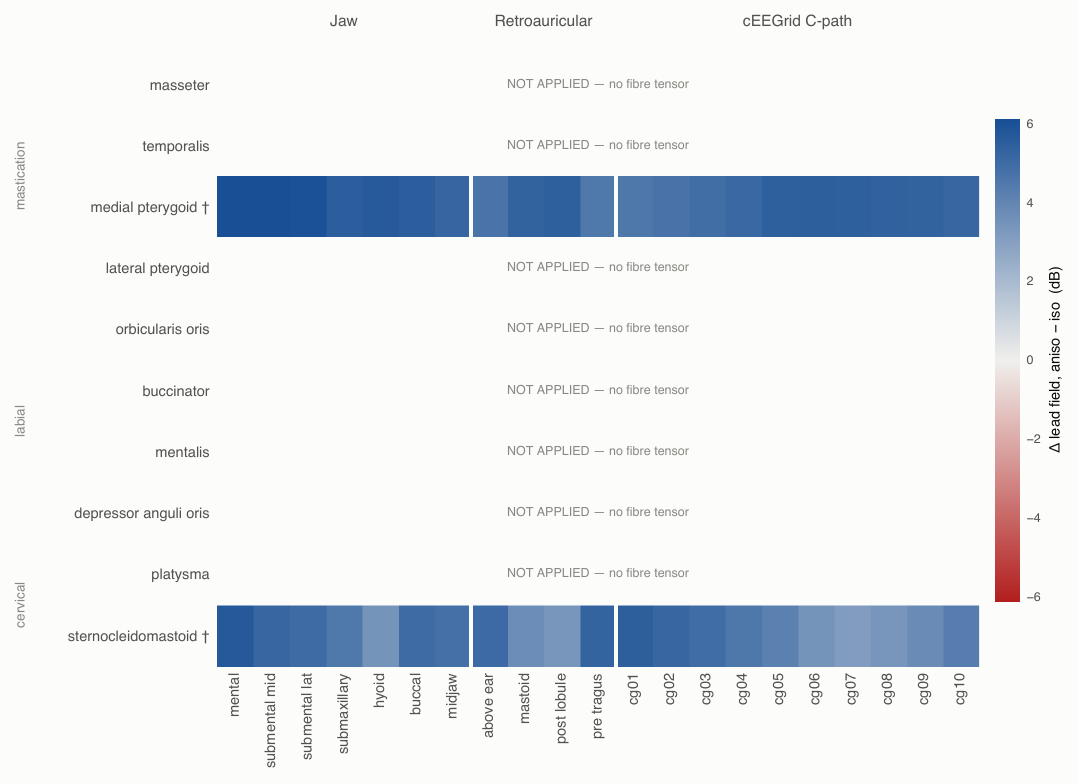}
\caption{Anisotropy robustness check. Change in lead field between the isotropic and anisotropic conditions, 20\textperiodcentered{}log\textsubscript{1}\textsubscript{0}(aniso/iso), per cell. The anisotropic condition is solved on the isotropic run's own mesh, so the two differ in conductivity alone and carry no electrode-realisation noise. A fibre tensor is applied to 2 of 10 muscles, sternocleidomastoid and medial pterygoid. Rows without a tensor are labelled not applied, not zero: they were never varied, so no null result was measured for them.}
\end{figure}

For the tensors tested at this resolution, modelling anisotropy changes
absolute lead-field values but has little effect on coupling \emph{ratios}
between electrode sites. This distinction matters when choosing which
model detail a device comparison requires.

\subsubsection{3.3 The tissue-conductivity contrast is a small term with a muscle-dependent sign}\label{the-tissue-conductivity-contrast-is-a-small-term-with-a-muscle-dependent-sign}

Solving the full montage twice on identical geometry, once with adipose at
0.025 S/m and once with both adipose compartments set to muscle conductivity,
attributes any difference to material properties alone, since source-to-electrode
distance is unchanged by construction. The second condition is a counterfactual
used to decompose mechanism; the gaps reported throughout this paper are the
first, because real anatomy contains adipose tissue.

We use this comparison to estimate how much the adipose conductivity
contrast contributes to the jaw advantage. An earlier version described the
comparison in terms of a retroauricular deficit; we now report the signed
jaw-versus-ear gap and its material contribution. The uniform and derived-field comparisons are distinguished in \S{}3.1; the conductivity decomposition here does not resolve their anatomical assumptions.

The contribution is not uniform in sign across muscles, because which electrode
is best differs by muscle and the shift is not uniform within a montage:

{\def\LTcaptype{none} 
\begin{longtable}[]{@{}
  >{\raggedright\arraybackslash}p{(\linewidth - 10\tabcolsep) * \real{0.2200}}
  >{\raggedright\arraybackslash}p{(\linewidth - 10\tabcolsep) * \real{0.1600}}
  >{\raggedright\arraybackslash}p{(\linewidth - 10\tabcolsep) * \real{0.1800}}
  >{\raggedright\arraybackslash}p{(\linewidth - 10\tabcolsep) * \real{0.1400}}
  >{\raggedright\arraybackslash}p{(\linewidth - 10\tabcolsep) * \real{0.3000}}@{}}
\toprule\noalign{}
Muscle & As modelled & Without contrast & Change & Contrast's role \\
\midrule\noalign{}
\endhead
\bottomrule\noalign{}
\endlastfoot
temporalis & \textminus{}3.801 & \textminus{}4.923 & \textminus{}1.121 & suppresses 29.5 \% \\
sternocleidomastoid & \textminus{}1.958 & \textminus{}1.547 & 0.411 & contributes 21.0 \% \\
lateral pterygoid & \textminus{}1.855 & \textminus{}1.532 & 0.323 & contributes 17.4 \% \\
medial pterygoid & 1.147 & 1.245 & 0.098 & suppresses 8.6 \% \\
masseter & 1.700 & 1.668 & \textminus{}0.032 & contributes 1.9 \% \\
orbicularis oris & 8.192 & 9.286 & 1.093 & suppresses 13.3 \% \\
platysma & 8.844 & 8.901 & 0.057 & suppresses 0.6 \% \\
buccinator & 10.033 & 9.848 & \textminus{}0.185 & contributes 1.8 \% \\
depressor anguli oris & 14.607 & 14.001 & \textminus{}0.606 & contributes 4.1 \% \\
mentalis & 21.945 & 19.082 & \textminus{}2.863 & contributes 13.0 \% \\
\end{longtable}
}

We report the changes per muscle. A population differential has no clean
definition under statistic A: the median change is +0.01 dB and conceals a sign spanning
\textminus{}2.86 to +1.09 dB, so a single percentage would conceal the differences between muscles.

For the labial group, where the jaw's advantage is largest, the contrast
accounts for a small and inconsistent fraction of it, and not in one direction:
the contrast contributes 2.86 dB to the mentalis gap and reduces
the orbicularis oris gap by 1.09 dB. For the four muscles
whose gaps sit closest to zero, the contrast is of the same order as the gap
itself, which is part of why those gaps do not resolve.

These results support geometry as the main source of the jaw advantage. Removing the
single largest conductivity contrast in the intervening tissue leaves every
montage assignment unchanged and moves no gap across zero. This is consistent
with \S{}3.5, where collapsing all non-muscle soft tissue to one conductivity also
preserves every assignment while moving magnitudes.

\subsubsection{3.4 Truncation sensitivity}\label{truncation-sensitivity}

The inferior truncation lies closer to the jaw montage than to the
retroauricular montage, so the cut plane could in principle inflate the jaw
side. Perpendicular clearance to the fitted plane places 2 jaw sites within
10 mm of it (\texttt{hyoid} 7.76 mm and \texttt{submental\_lat} 9.76 mm); the next nearest is
\texttt{submental\_mid} at 10.76 mm. The closest ear site, \texttt{cg09}, is 75.27 mm away.

Removing the near-cut sites moves the median gap from +4.68 dB to +4.52 dB, a
shift of \textminus{}0.17 dB. No muscle changes sign (0 of 10). The largest individual
movements are \texttt{medial\_pterygoid} \textminus{}0.88, \texttt{sternocleidomastoid} \textminus{}0.54 and
\texttt{platysma} \textminus{}0.37 dB. The reported advantage survives exclusion of the sites closest to
the truncation. The sensitivity reported here measures the effect of excluding
the most exposed sites. It does not bound the residual boundary artefact on the
sites that remain, which is addressed in \S{}4.7.

The 10 mm grouping above is descriptive. The reported subsets span every
admissible site set for any near-cut threshold between 9.757 mm and 15.264 mm, a
window of 5.507 mm, so no conclusion here depends on where in that range the
threshold is placed.

For muscles whose gaps do not change, the best jaw electrode is outside
the excluded set. Removing the near-cut sites therefore leaves their
maximum coupling unchanged.

\begin{center}\rule{0.5\linewidth}{0.5pt}\end{center}

\subsubsection{3.5 A homogeneous conductor reaches the same verdicts}\label{a-homogeneous-conductor-reaches-the-same-verdicts}

A homogeneous soft-tissue conductor reproduces every montage assignment.
Setting skin, adipose and the non-muscle soft tissues to a single conductivity,
with geometry, electrodes and sources held exactly fixed, changes no muscle's
montage preference. Eight of ten gap magnitudes move by more than the 0.27 dB
floor, with a median shift of 0.482 dB and a maximum of 3.271 dB (mentalis).

The direction is not uniform. Temporalis's retroauricular advantage \emph{grows}
under the homogeneous conductor, from \textminus{}2.571 to \textminus{}3.724 dB at the pre-registered
cluster {[}\texttt{results/04r\_homog\_cluster.csv}{]}, so tissue detail reduces the ear advantage under this
uniform-orientation comparison.
Sternocleidomastoid and lateral pterygoid move the other way.

In this comparison, the simplified conductor preserves montage assignments
but changes the coupling gaps. Tissue detail therefore matters for a
design comparison that uses gap magnitudes in decibels.

\subsubsection{3.6 Targeted placement improves lateral-pterygoid coupling}\label{placement-by-anatomical-target-outperforms-density}

The cluster chosen by anatomical target improves coupling for lateral
pterygoid relative to arbitrary four-site placement. The
four-site retroauricular cluster, above the ear, over the mastoid, behind and
below the lobule, and anterior to the tragus, was specified by anatomical
target in the project repository before any solve was run. Compared against
the median of random four-site draws from the same fourteen candidates, it is
1.07 dB better for lateral pterygoid (\textminus{}1.564 against \textminus{}0.498) and equivalent
for sternocleidomastoid (\textminus{}0.973 against \textminus{}0.979)
{[}\texttt{results/04h\_matched\_counts.csv}{]}.

Neither of the two sites that won the unmatched argmax for temporalis and
sternocleidomastoid is in that cluster, so the best sites in the full candidate set do not coincide with the
anatomically chosen four-site montage. An argmax over fourteen sites and
a comparison at fixed electrode count answer different design questions.
For a device with four contacts, the matched-count comparison measures
the benefit of selecting their locations.

\subsection{4. Discussion}\label{discussion}

\subsubsection{4.1 Montage preferences depend on the source model}\label{the-montages-are-not-complementary}

Five articulators consistently favour the jaw across the tested orientation and electrode-subset controls. Ear-side preferences are less stable across modelling choices. The derived temporalis interval crosses zero, whereas the derived lateral-pterygoid interval excludes it on the ear side (\S{}3.1). That result permits a conditional coupling advantage for an ear montage, but not a claim of general montage complementarity or a decoding advantage.

The retroauricular montage is 8.11 to 22.40 dB less sensitive to the labial
group, without a reliable compensating advantage for another modelled muscle.
Proximity to a bony attachment identifies competitive retroauricular sites and
predicts their ordering, but does not establish an advantage over jaw sites.

The temporalis result depends on how source orientation is modelled. Under a uniform orientation sweep it favours the ear by 2.57 dB with an
interval excluding zero. Under the fibre field derived from the label volume it
favours the ear by 1.147 dB with an interval of {[}\textminus{}1.453, +5.458{]}, and does not
resolve. The two treatments respectively assume uniformly distributed source
orientations and derive orientation from the segmented muscle's attachment. That difference also changes what the
accompanying intervals measure. Under the uniform sweep, orientation is a
sampled dimension and the interval averages over 200 directions within each
draw; under the derived field there is no orientation dimension to average over,
and the interval reflects site selection alone. The two intervals answer
different questions and should not be read as one quantity under two fibre
models. Deriving directions from a segmented attachment still imposes an anatomical model; it does not eliminate orientation uncertainty. Both temporalis and lateral pterygoid have unresolved construction limitations (\S{}4.7). The conclusion supported across the reported treatments is the five-muscle jaw advantage, not an unconditional absence of ear-side advantages. The companion experiment in \S{}4.6 is intended to test the conditional coupling predictions.

\subsubsection{4.2 An anatomical prediction that did not survive}\label{an-anatomical-prediction-that-did-not-survive}

The prediction recorded before solving (\S{}2.8) named temporalis and
sternocleidomastoid as candidates for stronger retroauricular coupling.
Temporalis originates in the temporal fossa beneath the superior cEEGrid row;
sternocleidomastoid inserts on the mastoid process.

The uncontrolled comparison agreed for those two muscles and also showed an
ear advantage for lateral pterygoid, which the recorded prediction had not
named. Its insertion at the mandibular condyle is adjacent to the temporal
bone's mandibular fossa. These three muscles, and only these three, showed a
retroauricular advantage in that comparison. After matching electrode counts and applying the derived temporalis fibre
field, neither predicted muscle retains a robust ear advantage. Lateral pterygoid, which the prediction had not named, retains an ear-side interval under its derived fan (\S{}3.1, \S{}4.7).

We therefore report the pre-solve prediction as failed. Attachment proximity
predicts which retroauricular sites are competitive and their ordering: the
three near-temporal muscles have the gaps closest to zero. It does not reliably
predict an ear advantage. Current paths depend on the volume conductor, so an
attachment next to an electrode need not make that electrode more sensitive to
the muscle's source activity.

\subsubsection{4.3 Geometry and intervening tissue contribute differently}\label{the-mechanism-is-distance-not-intervening-tissue}

For the labial group the ear's deficit is predominantly geometric, though the
material contribution varies widely within the group: the adipose--muscle
conductivity contrast accounts for 0.6 per cent of the gap for platysma and
13.3 per cent for orbicularis oris. The remainder is attributed to geometry, including source-to-electrode
distance. Figure 5 shows the attenuation-against-distance relation. The share is largest where the gap is
smallest, and that is mostly a denominator effect rather than a difference in
the tissue's role. The absolute contribution of the contrast is of comparable
size across muscles, 0.03 to 2.86 dB, while the gaps it is measured against
span an order of magnitude. The material share is 21.0 per cent for
sternocleidomastoid, 17.4 per cent for lateral pterygoid, and 29.5 per cent
against the gap for temporalis. It falls to 0.6 per cent for platysma, whose
jaw advantage is much larger. These percentages compare the same material contribution against different
gap magnitudes.

\begin{figure}[htbp]
\centering
\includegraphics[width=\linewidth,height=0.42\textheight,keepaspectratio]{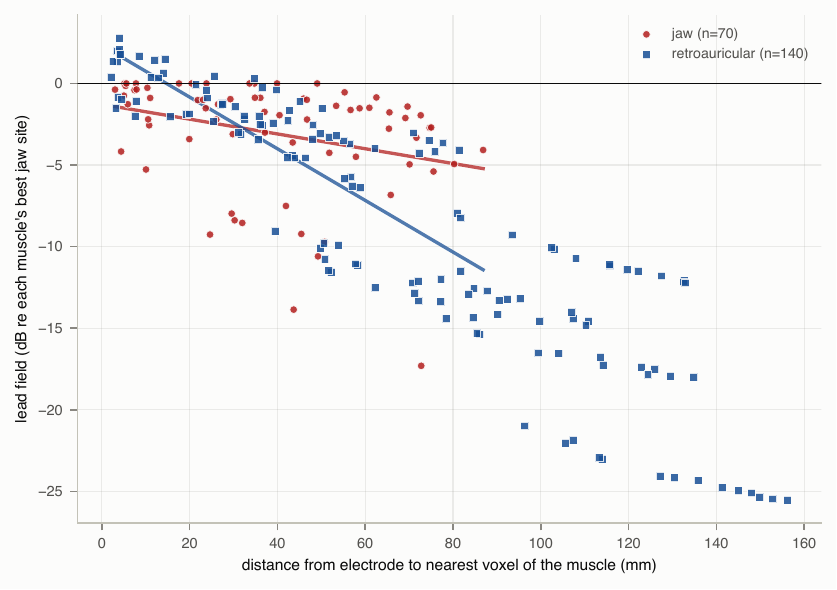}
\caption{Attenuation against source-electrode distance. One point per (electrode, muscle) pair: straight-line distance from the electrode to the nearest voxel of that muscle compartment against the lead field in dB relative to that muscle's best jaw site, isotropic condition, truncated mesh. The two montages do not lie on the same attenuation curve. Fitted over the 3 to 87 mm range where both montages have electrodes, the retroauricular sites fall off at \textminus{}0.158 dB/mm against the jaw's \textminus{}0.045 dB/mm, a factor of 3.5 [results/04s\_fig3\_slopes.csv]. Within these modelled electrode-muscle pairs, the retroauricular fit has a steeper distance-associated decline. This descriptive comparison is consistent with a geometric contribution to the labial-group deficit in \S{}4.3; it does not isolate the effect of moving an individual electrode. Lines are least-squares fits over the overlapping range only, so the comparison is made where both montages have data, not extrapolated across the ear's longer reach.}
\end{figure}

Limb studies cannot make this separation, because adding a fat layer changes
material properties and source-to-electrode distance together (Kuiken et al.
2003). A labelled head model can, because conductivity is changed with geometry
held exactly fixed. In this model the comparison requires one additional solve.

The material share is not a dose response. Across the muscles for which a layer
profile exists, it correlates \emph{negatively} with the adipose fraction of the
muscle-to-skin path (Spearman $\rho$ = \textminus{}0.955, p = 0.001, n = 7): platysma sits at
0.650 fat fraction and 0.6 per cent material share, while sternocleidomastoid
sits at 0.225 and 21 per cent. At n = 7 a single muscle could carry that
relationship, so it was tested: dropping each muscle in turn leaves $\rho$ between
\textminus{}0.928 and \textminus{}0.986 and p below 0.01 in all seven cases
{[}\texttt{results/04t\_correlation\_robustness.csv}{]}; the seven muscles and that
relationship are plotted in Figure 6. The negative association is consistent with cancellation between the two
montages' conductivity responses. The reported share is \textbar $\Delta$gap\textbar{} / \textbar gap\textbar, and
a muscle embedded uniformly in fat has both the jaw and the ear route shifted
together, so the change subtracts out of the ratio. The quantity that would
track positively is the \emph{difference} between how the two routes traverse fat,
which this study does not form. A uniform magnitude offset likewise cancels from these ratios. We had not
anticipated this effect.

\begin{figure}[htbp]
\centering
\includegraphics[width=\linewidth,height=0.42\textheight,keepaspectratio]{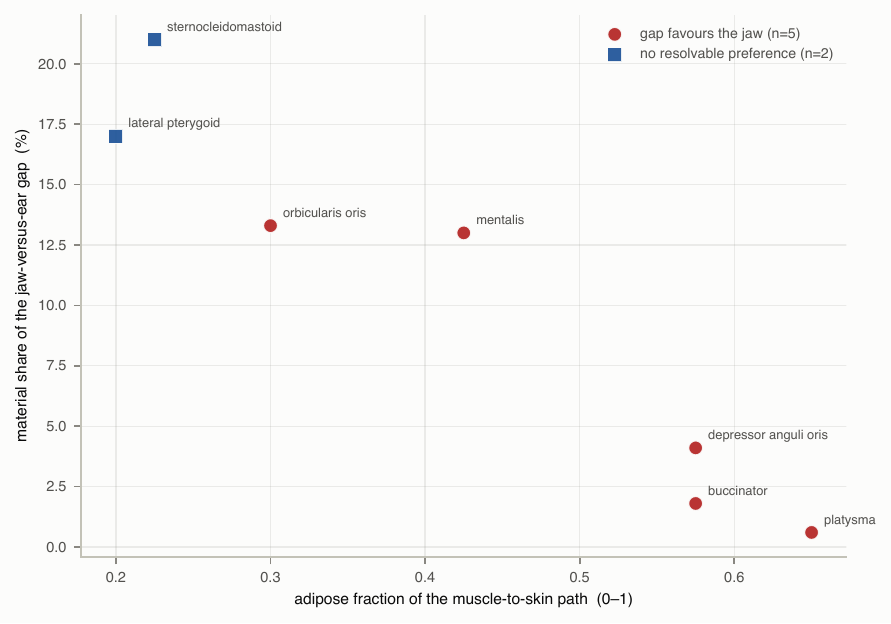}
\caption{Material share falls as the adipose path grows. For each of the seven muscles with a layer profile, the adipose fraction of the muscle-to-skin path against the material share of that muscle's jaw-versus-ear gap. The relationship is negative (Spearman rho = -0.955, p = 0.001, n = 7): greater adipose fraction is associated with a smaller material share. One possible explanation is cancellation: if the jaw and ear routes undergo similar conductivity-dependent changes, their ratio changes little. This association does not independently establish that mechanism. Leave-one-out leaves rho between -0.928 and -0.986 with p below 0.01 in every case, so no single muscle carries it. Points are marked by whether the muscle's gap resolves to the jaw or reaches no resolvable preference. No trend line is drawn: these are seven independent muscles, not a series, and Spearman is a rank statistic that claims no functional form.}
\end{figure}

Applied to the three muscles whose gaps come closest to zero, temporalis,
sternocleidomastoid and lateral pterygoid, the same decomposition returns a
modest term whose sign varies between them.

For temporalis the contrast acts against the gap, not for it: removing it
would enlarge the gap from \textminus{}3.801 to \textminus{}4.923 dB. In this comparison, the tissue contrast reduces the magnitude of the
temporalis gap.
It does not change the verdict, which is set by the matched\textminus{}count interval and
not by this term. For sternocleidomastoid and lateral pterygoid the contrast acts
with the gap, supplying 21.0 and 17.4 per cent of it respectively.

Across all ten muscles the decomposition has the same shape, drawn in Figure 7: a
geometric term that sets the size of the gap and a material term that trims it.
The material term never exceeds 2.9 dB in magnitude, against geometric terms
reaching 19.1 dB, and it changes sign. It moves five muscles toward the jaw and
five toward the ear. That sign change is the reason no single material share is
quoted for the set, and the reason the shares above are reported muscle by
muscle.

\begin{figure}[htbp]
\centering
\includegraphics[width=\linewidth,height=0.42\textheight,keepaspectratio]{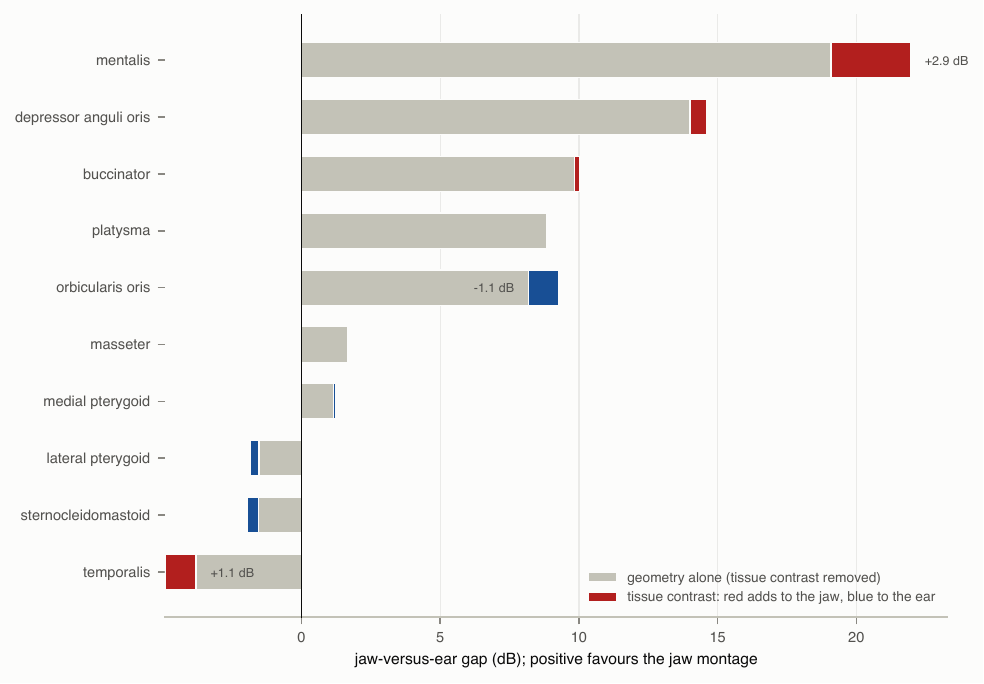}
\caption{Geometry sets the gap, tissue only trims it. Each muscle's jaw-versus-ear gap split into the part that survives when the adipose-muscle conductivity contrast is removed, with geometry held exactly fixed, and the part that does not. Positive is a jaw advantage. The geometric term reaches 19.1 dB; the material term never exceeds 2.9 dB and changes sign, moving five muscles toward the jaw and five toward the ear. A single material share for the set would average that sign change away, which is why the text reports shares muscle by muscle.}
\end{figure}

The decomposition suggests a hypothesis about transfer between subjects.
A margin dominated by geometry depends on anatomical variation; one with a
larger material contribution also depends more strongly on adipose distribution.
We have not measured either source of between-subject variation here. The two classes are therefore defined by material
share, not by which montage wins. A margin with a small share should transfer
more readily than one with a large share: 0.6 to 13.3 per cent across the labial
group, against 21 and 17 per cent for sternocleidomastoid and lateral pterygoid.
Temporalis, masseter and medial pterygoid are not classified, because no layer
profile exists for them and their share is unmeasured. This is a prediction from the decomposition that requires testing in a second
anatomy.

Figure 8 puts the mechanism on the anatomy itself. Taking one representative
electrode from each montage, mental for the jaw and mastoid for the ear, and
mapping the ratio of their two sensitivity fields on a plane that contains both
electrodes and the orbicularis oris centroid, the head separates into a jaw
territory and an ear territory with a crossover running between them.
Orbicularis oris sits 45 mm from the jaw electrode and 141 mm from the ear
electrode, and the jaw is 13.5 dB louder there; temporalis sits on the far side
of the crossover, 116 mm from the ear electrode against 155 mm from the jaw, and
the ear is 4.4 dB louder. The second panel shows the two electrodes falling on
one decay curve, not two. Pooled over the twenty compartment-and-electrode
pairs, log \textbar E\textbar{} against log distance gives Spearman rho = -0.961 (p = 2e-11) with
a slope of -1.42. For nine of the ten muscles the closer electrode is the louder
one; the exception is masseter, where the ear electrode is nearer by 1.9 mm and
the jaw is nonetheless 4.1 dB louder, and it is the only muscle in the set whose
ordering is decided by something other than distance. The figure plots \textbar E\textbar{} for single electrodes. Montage verdicts
use the projected lead fields and remain those reported in Table 4.

\begin{figure}[htbp]
\centering
\includegraphics[width=\linewidth,height=0.42\textheight,keepaspectratio]{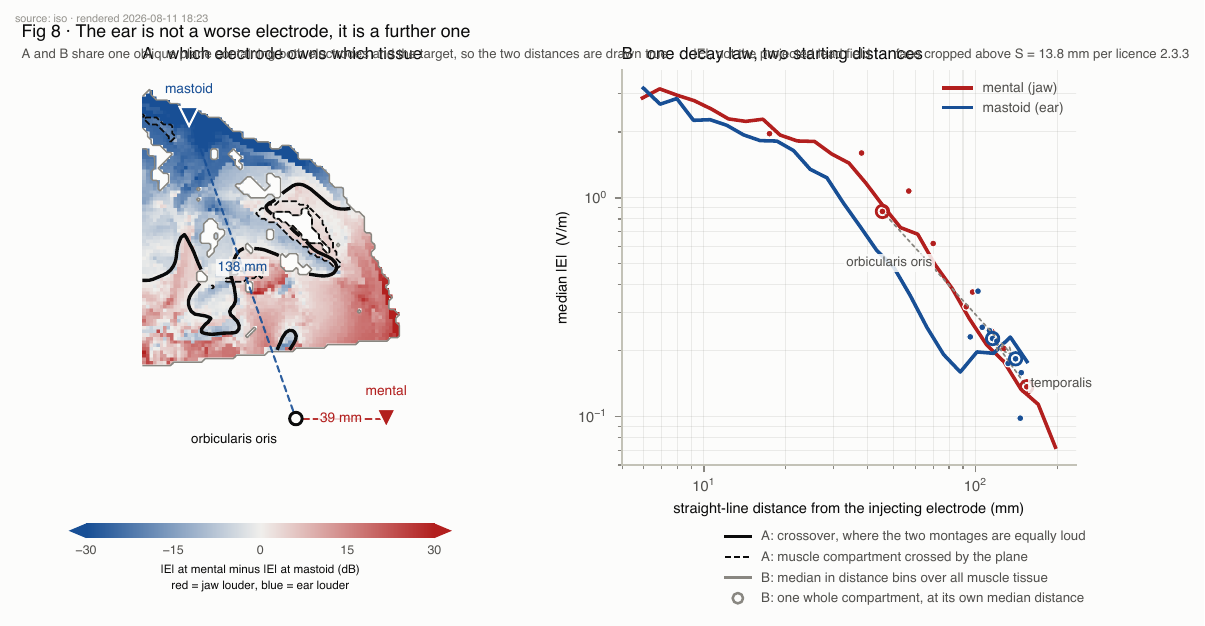}
\caption{The gap follows distance. \emph{Left:} the ratio of two sensitivity fields, one from a jaw electrode (mental) and one from a retroauricular electrode (mastoid), on a plane containing both electrodes and the orbicularis oris centroid, so the two distances are drawn true, not projected. Red is a jaw advantage, blue an ear advantage, and the heavy line is the crossover where the two are equally sensitive. \emph{Right:} median |E| against straight-line distance from the injecting electrode, in distance bins over all muscle tissue, with each segmented compartment marked at its own median distance. The two electrodes lie on one decay curve, not two: pooled over the twenty compartment-and-electrode pairs, log |E| against log distance gives Spearman rho = -0.961 (p = 2e-11), slope -1.42. For nine of ten muscles the closer electrode is the louder one; masseter is the exception. |E| is the field magnitude for a single electrode, the upper bound over source orientations, not the projected lead field for a montage, so this figure carries the mechanism and not the verdicts. This retained magnitude-only figure does not divide each site's field by delivered current; its ratio, crossover and fitted quantities retain that calibration dependence (Table 3, row 8). They are not current-normalized coupling estimates. The plane is cropped above the orbital rim under MIDA licence clause 2.3.3.}
\end{figure}

\subsubsection{4.4 What this licenses for device design}\label{what-this-licenses-for-device-design}

A device designer considering retroauricular electrodes can use the model to
estimate the sensitivity cost for particular muscles. The labial group loses
8.1 to 22.4 dB relative to the jaw montage. The derived lateral-pterygoid interval supports an ear-side advantage at the fixed jaw basis, but that preference is not established across fibre models. The labial gaps quantify
part of the signal cost of choosing an ear-worn form factor for wearability.

The uniform-sweep comparisons and the derived temporalis comparison depend on electrode selection (\S{}3.1). The derived lateral-pterygoid interval is wholly ear-side, but its anatomical construction remains unvalidated. These are different reasons for uncertainty in choosing a four-contact device. Comparisons that search
unequal numbers of sites can overstate the benefit of the more densely sampled
montage.

Placement within the ear array still matters. A four-site cluster chosen by
anatomical target improves lateral-pterygoid coupling relative to the median
of random four-site draws from the same candidates, while giving essentially
the same coupling for sternocleidomastoid (\S{}3.6). Devices with few contacts may therefore benefit from
selecting sites for their intended muscle targets.

This study does not identify an optimal montage. Selecting the best four of the twenty-one modelled positions that belong to a montage is an argmax over densely spaced candidates, which is the
operation \S{}4.8 shows to be untrustworthy: ten of the fourteen retroauricular
positions lie on a cEEGrid path at 12 to 18 mm spacing, so a search across them
rewards how many were solved. The candidate sampling therefore does not support an optimal-montage
recommendation.

What can be reported without such a search is the strongest single site for each
articulator, with its margin to the runner-up, given as Table 5. The margins are
small, 0.03 to 1.02 dB, and four of the ten sit at or below the 0.65 dB upper
bound of the meshing floor's 95 \% CI, so for mentalis, depressor anguli oris,
platysma and sternocleidomastoid this model does not resolve which single
electrode is best. No muscle's runner-up sits in the other montage, though: all
ten second-best sites share a montage with their best. The best single site is thus less well resolved than the montage membership
of the two strongest sites. This checks an assumption underlying the montage
comparisons.

The strongest site is a jaw site for seven of the ten articulators. The three
served best by a retroauricular site, temporalis, lateral pterygoid and
sternocleidomastoid, are the three whose montage-level advantages do not survive
the controls (\S{}4.8). A muscle's strongest single electrode can belong to the ear montage even
when a four-site ear montage has no robust advantage over a four-site jaw
montage.

A montage recommendation would need two things this study does not supply.
Coupling strength is not decoding accuracy, and two sites coupling strongly to
one muscle carry less joint information than their individual couplings suggest;
what a decoder consumes is channel independence. It would also need the
articulators this model cannot separate, since the tongue and suprahyoid
compartments are pooled in MIDA (\S{}4.7) and carry a large share of articulation,
so no placement statement derived here covers them.

\subsubsection{4.5 Contamination, described muscle by muscle}\label{contamination-described-muscle-by-muscle}

The EEG literature has documented mastoid and retroauricular electromyographic
contamination for decades as a nuisance to be suppressed {[}Goncharova et al.
2003; Yao et al.~2019{]}. This model identifies possible muscle contributions to that contamination. The
three compartments a retroauricular electrode couples to most strongly, relative
to the canonical jaw montage (Figure 2), are temporalis, sternocleidomastoid
and lateral pterygoid, and their best positions differ, so contamination at
\texttt{cg01} is not the same mixture as contamination at \texttt{cg08}. The pooled
suprahyoid compartment contributes to that mixture as well; its field under
the retroauricular montage is shown in Figure 9.

\begin{figure}[htbp]
\centering
\includegraphics[width=\linewidth,height=0.42\textheight,keepaspectratio]{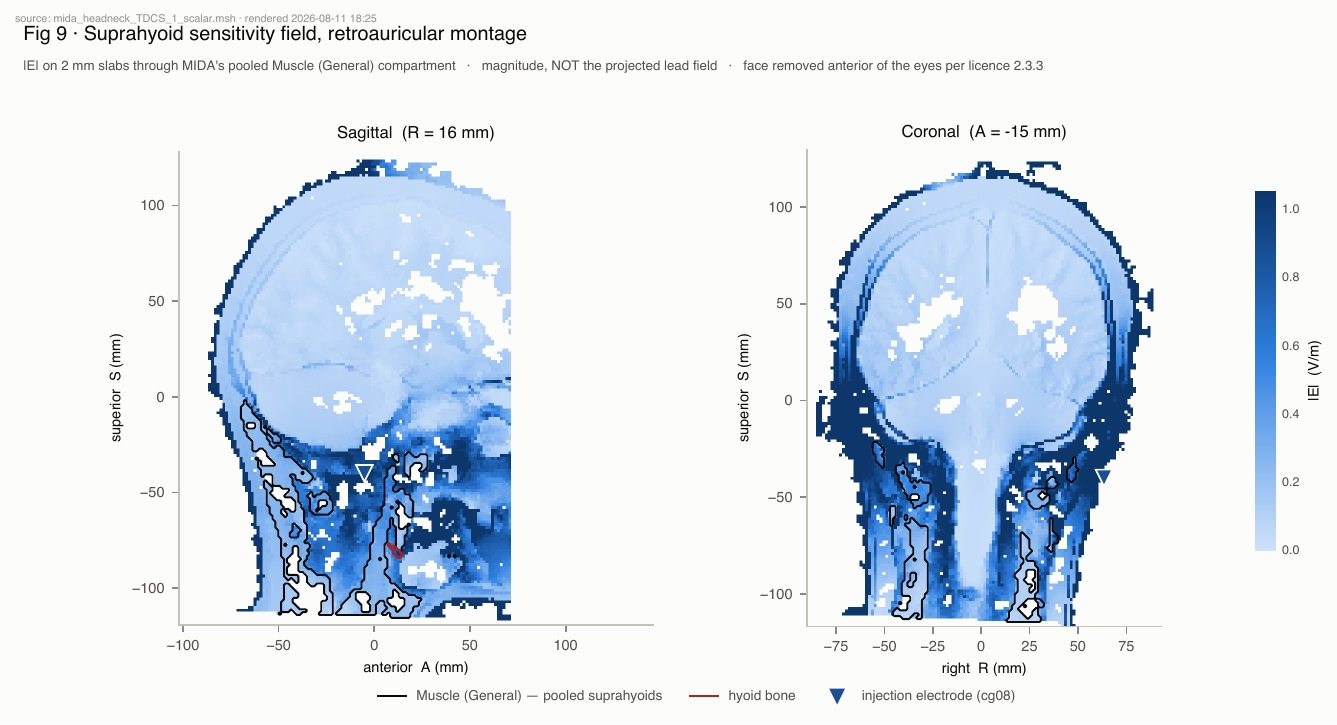}
\caption{Suprahyoid sensitivity field. Sagittal and coronal slices of |E| through the pooled Muscle (General) compartment for the retroauricular montage, with the mastoid notch, hyoid greater horn and the corridor between them overlaid. Reported as a field and not as per-muscle values because MIDA does not segment the suprahyoid group.}
\end{figure}

These differences may inform artifact rejection. A spatial filter could
account for the muscles most strongly coupled to each contact, rather than
treating retroauricular electromyography as a single component.

\subsubsection{4.6 A specific prediction for a companion experiment}\label{a-specific-prediction-for-a-companion-experiment}

A companion experiment would record four jaw and four retroauricular channels
simultaneously. The first public version of this paper predicted that no articulator would carry
more information in the ear channels. We withdraw that inference from coupling:
the forward model does not determine decoder performance or channel independence.
The experiment should measure muscle-specific coupling as well as decoding
performance, because the model predicts electrical sensitivity rather than the
information a classifier can extract from utterances.

The clearest prediction is lower ear coupling for the five labial and chin
muscles under comparable source activation. For temporalis,
sternocleidomastoid and lateral pterygoid, the comparison depends on electrode
selection and source orientation. The experiment should specify those choices
in advance and compare measured coupling with the corresponding model result,
including the alternative lateral-pterygoid fan disclosed in \S{}4.7.

A reproducible disagreement with the matched coupling prediction would challenge
the model under those conditions. An advantage in decoding accuracy alone would
require further explanation: source activation, noise, correlations between
channels, and mechanical or acoustic coupling are not determined by the
electrical sensitivity comparison.

\clearpage

\subsubsection{4.7 Limitations}\label{limitations}

\textbf{Mesh realisation is wider than four of the reported margins.} Term 10 of the
error budget is a rebuild of the same nominal mesh, and it moves per-muscle gaps
by up to 1.554 dB. Four verdicts have envelopes comparable to that: masseter
(+1.20 to +2.22), medial pterygoid (+0.86 to +1.33), sternocleidomastoid (\textminus{}2.53
to \textminus{}0.97) and lateral pterygoid (\textminus{}3.61 to \textminus{}1.53). Medial pterygoid's envelope
sits inside it entirely, so that verdict in particular should not be relied on
until the term is characterised over several rebuilds. The five labial verdicts
span 8.11 to 22.40 dB and are unaffected, so the study's headline result does
not depend on this term. It is listed as unquantified rather than bounded
because one realisation pair is not a distribution. The rebuild comparison is
\texttt{results/04w\_control\_mesh\_realisation.csv}, the refinement comparison
\texttt{results/04w\_mesh\_convergence.csv}, and the estimator check
\texttt{results/04x\_estimator\_stability.csv}.

\textbf{Ten of eighteen muscles are modelled, and the two carrying the strongest
version of the anatomical argument are not among them.} MIDA does not
individually segment the suprahyoid group or the tongue. Posterior digastric and
stylohyoid, the two muscles that anchor at the mastoid notch and styloid
process, and that motivated the retroauricular hypothesis in the first place,
are therefore absent from the per-muscle comparison. The per-muscle comparison therefore excludes the muscles most directly
motivating the retroauricular hypothesis. The spatial sensitivity field over
the pooled compartments provides partial information but cannot recover their
individual couplings.

MIDA is a single subject, and between-subject variance in muscle geometry,
adipose thickness and pinna position cannot be estimated from one head. The adipose decomposition suggests that this uncertainty may differ by
muscle. The labial group and temporalis are carried by geometry,
which is comparatively conserved between individuals; sternocleidomastoid and
lateral pterygoid draw 21.0 and 17.4 per cent of their gap from the conductivity
contrast, so of the three muscles nearest zero they are the two whose position
is most dependent on tissue properties, not on geometry, and the two whose gaps
should be expected to move most with subject adiposity. This suggests differential generalisation, but testing it requires additional
anatomies.

\textbf{The inferior boundary is an unquantified limitation whose bias runs against
the ear.} A neck-extended mesh was built specifically to measure it and did not
conserve charge, so the pre-committed decision rule was recorded unexecuted, not
applied or revised. What can be said is \S{}3.4: excluding the two near-cut jaw
sites moves the median gap by only \textminus{}0.17 dB and flips no signs, and seven of ten
muscles are immune by construction. The residual bias remains unquantified, and its expected direction favours
the paper's headline comparison.

\textbf{Static geometry.} The model is solved on a single anatomical configuration.
Articulation moves the tongue, opens and closes the oral cavity and alters the
airway, all of which change the volume conductor during the task being measured.
These effects are unquantified. Jaw electrodes are closer to the moving
structures than ear electrodes.

\textbf{Quasi-static assumption}, standard at surface electromyography frequencies
and stated for completeness.

Fibre orientation is bounded, not known (\S{}2.5). A fibre tensor reaches two of
ten segmented muscles, and rows without one are reported as not applied and not
as zero change.

\textbf{Anatomically derived directions remain model assumptions.} The construction in \S{}2.5.1 points voxels toward a mandibular insertion. MIDA label 63 combines temporalis with temporoparietalis, which does not share that mandibular insertion. Applying one fan to the whole label therefore assigns the latter voxels an inappropriate attachment. Their contribution and its effect on the gap are unquantified. The lateral-pterygoid construction also requires validation because a common insertion centroid may not represent its two heads. Neither uncertainty has been resolved here.

The construction requires a discrete bony insertion for voxels to point at, so
the remainder do not form a single pending set. Sternocleidomastoid, medial
pterygoid and mentalis are strap-like with fibres along the compartment long
axis, and admit it directly. Masseter, lateral pterygoid and depressor anguli
oris are multi-part or converging, superficial and deep layers at different
angles, two heads, and a converging triangular sheet respectively. The
per-voxel fan was pre-registered and run for the first two: masseter voids the
construction, its insertion patch spanning 66 to 89 per cent of the
compartment on every axis, so voxel-to-insertion is not a fibre model there;
lateral pterygoid gives a derived interval of {[}\textminus{}5.327, \textminus{}1.115{]},
wholly on the ear side, whereas the main four-site uniform-sweep interval
crosses zero. The pre-registration named this fan as an appropriate treatment and specified that its result governs where the construction is valid (R3). Its post-run Amendment 1 introduced the rule that rejects masseter; lateral pterygoid passed those tests. The later R5 entry records disclosure without a title or abstract change. That history does not support excluding the lateral-pterygoid result while invoking R3 for temporalis. This revision reports both derived intervals and withdraws the unconditional no-ear-advantage title.

The stored extension flag for whether R2 fired uses a uniform jaw-subset envelope, whereas the rule specifies an exact ear-subset comparison. No exact uniform-sweep ear-subset interval for lateral pterygoid is supplied. We therefore report the available intervals and their bases rather than claiming that R2 either fired or did not fire. No pre-registration entry or result is rewritten. Depressor anguli oris remains untested. Orbicularis oris admits it in no form: it is a sphincter whose
fibres run in a ring, with no bony insertion, so a principal axis would not represent its fibre geometry. Buccinator blends into that sphincter at the
modiolus and inherits the same problem. Platysma is a broad sheet.

The treatment that changed the temporalis result is applicable to some
compartments but cannot represent the fibre geometry of all nine remaining
muscles.

\textbf{One jaw site is withheld, and its omission runs against the reported jaw
advantage.} \texttt{throat\_scm} carries no coordinate because MIDA's
sternocleidomastoid is truncated by the inferior cut plane, which biases its
centroid posteriorly, so no defensible automatic placement exists (\S{}2.3). It is
the jaw site nearest the ear montage, so a jaw montage carrying it would be
compared from a position closer to the retroauricular sites than any jaw site
actually used. The direction of that omission is toward a smaller jaw advantage
than is reported.

\subsubsection{4.8 How the retroauricular advantage dissolved}\label{how-the-retroauricular-advantage-dissolved}

The intermediate analyses showed a retroauricular advantage that appeared in
the first public version of this paper. The controls below progressively changed the temporalis estimate;
the final interval spans zero. The sequence describes the calculations; it does not establish when each modelling choice was made.

{\def\LTcaptype{none} 
\begin{longtable}[]{@{}
  >{\raggedright\arraybackslash}p{(\linewidth - 2\tabcolsep) * \real{0.5000}}
  >{\raggedright\arraybackslash}p{(\linewidth - 2\tabcolsep) * \real{0.5000}}@{}}
\toprule\noalign{}
\begin{minipage}[b]{\linewidth}\raggedright
Stage
\end{minipage} & \begin{minipage}[b]{\linewidth}\raggedright
Temporalis gap
\end{minipage} \\
\midrule\noalign{}
\endhead
\bottomrule\noalign{}
\endlastfoot
uncorrected field magnitude, best of 14 ear sites & \textminus{}3.92 dB \\
projected onto source orientation & \textminus{}3.31 dB \\
matched electrode counts, four sites each & \textminus{}2.57 dB, interval {[}\textminus{}3.31, \textminus{}0.03{]} \\
derived per-voxel fibre field & \textbf{\textminus{}1.15 dB, interval {[}\textminus{}1.45, +5.46{]} spans zero} \\
\end{longtable}
}

These stages do not isolate three independent effects. The first uses uncorrected field magnitudes; the projected comparisons use per-site current normalization (\S{}2.4). The first shift therefore combines projection with normalization. Reporting the field magnitude instead of the lead field
projected onto a source orientation overstates coupling, because the magnitude
is the maximum over orientations. Comparing the best of fourteen candidate sites
against the best of four rewards electrode density, not placement. And assuming
a fibre direction rather than deriving one from the label volume permitted a
claim the derived field does not support.

Each treatment can be defensible in a forward model, but the choice affects
the comparison. The successive changes each shift the estimate by around a
decibel. Together they change an apparent 3.92 dB ear advantage into a smaller
estimate whose interval spans zero.

Figure 10 draws that cascade with its intervals, and shows what the final one is
made of. The ear pool holds 14 candidate sites and a montage takes 4, so there
are exactly 1001 possible retroauricular montages and all of them are
enumerated: 50.5 per cent favour the ear and 49.5 per cent favour the jaw. The sign of the final comparison therefore depends on which four sites are
selected.

\begin{figure}[htbp]
\centering
\includegraphics[width=\linewidth,height=0.42\textheight,keepaspectratio]{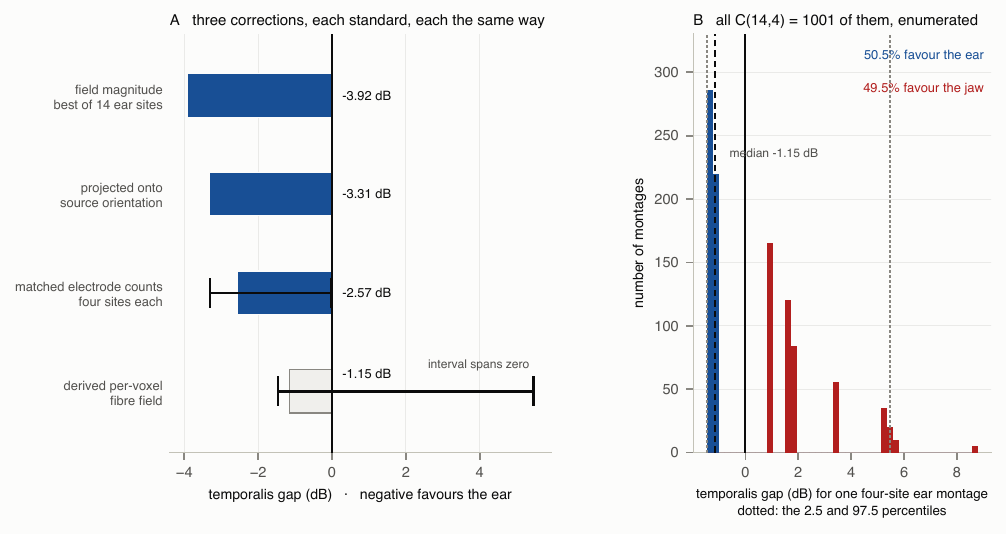}
\caption{How the retroauricular advantage dissolved. \emph{Left:} the temporalis gap at four stages, each read from the file that reproduces it: field magnitude against the best of fourteen ear sites, then projected onto source orientation, then at matched electrode counts, then under the fibre field derived from the label volume. Rules are matched-count intervals where one exists. The first-stage magnitude is uncorrected for delivered current; later projected stages are normalized. The first shift is therefore not a pure orientation effect. Only the last displayed interval spans zero. The order of these calculations is not evidence that every modelling choice preceded inspection of its result. \emph{Right:} what that last interval is made of. The ear pool holds 14 sites and a montage takes 4, giving exactly C(14,4) = 1001 possible retroauricular montages, every one of which is enumerated here, so the histogram is the population, not a sample of it. Dotted lines are the 2.5 and 97.5 percentiles quoted as the interval; 28.6 per cent of montages attain the lower bound exactly, which is why that bound coincides with the floor.}
\end{figure}

The direction of the successive shifts does not independently validate the
final result. Removing optimistic assumptions can produce a monotone sequence
by construction. Our reason for preferring the final treatment is that it
replaces an assumed orientation distribution with a field derived from anatomy,
as specified by the derived-field rule in the repository's fibre-fan
pre-registration. That rule also applies to the lateral-pterygoid construction, whose derived interval is reported in \S{}3.1. Neither construction is validated by the direction in which it moves the estimate.

The intermediate results were internally consistent, exceeded their
measurement floors and agreed with the a-priori anatomical prediction. The
three apparent ear advantages occurred in the three muscles attaching at or
near the temporal bone. That agreement made the result plausible, but did not
establish its robustness.

Figure 11 retains the uniform-orientation comparison and the temporalis refinement from the earlier version. It is not a complete summary of the derived-field results: lateral pterygoid also has a derived comparison (\S{}3.1). The graphical abstract displays both derived intervals. Figure 11's inherited title refers only to the refinement drawn there, not to all fibre-model dependence in this paper.

\begin{figure}[htbp]
\centering
\includegraphics[width=\linewidth,height=0.42\textheight,keepaspectratio]{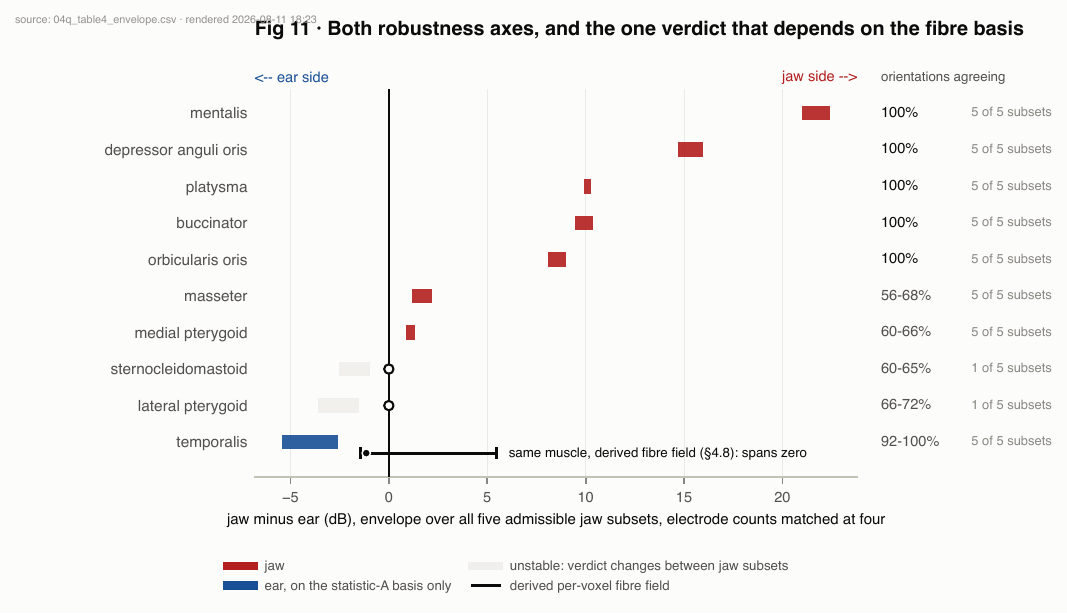}
\caption{Uniform-sweep robustness axes with the temporalis refinement. Table 4 drawn: one bar per muscle spanning the envelope over all five admissible jaw subsets, with electrode counts matched at four, coloured by verdict, against the orientation-agreement range on the right. Five muscles sit entirely on the jaw side on both axes. Two are unstable: their site-robustness verdicts vary between jaw subsets, although the plotted gap envelopes remain on the ear side. Temporalis sits entirely on the ear side on the statistic-A basis used for Table 4, which takes the median over a uniform 200-direction orientation sweep. The black rule beneath it is the derived-field interval across all 1001 four-site ear subsets at the fixed published jaw basis; it spans zero. It varies ear sites, unlike the coloured jaw-subset envelopes. Only temporalis carries a second rule in this retained figure; the derived lateral-pterygoid interval is reported in \S{}3.1 and in the graphical abstract. The figure's inherited title is limited to the comparison shown. Section 4.8 gives the argument for preferring the temporalis derived field.}
\end{figure}

\begin{center}\rule{0.5\linewidth}{0.5pt}\end{center}

\clearpage

\subsection{Tables}\label{tables}

\textbf{Table 1, Tissue conductivities.} All 116 MIDA labels from \texttt{results/01\_table1\_conductivities.csv}, reproduced below. S = SimNIBS 4.6 head-modelling default; I = IT'IS LF v4.2 (DOI 10.13099/VIP21000-04-2); * = judgement assignment. IT'IS values are evaluated at 100 Hz; defaults and numerical air conductivity are not measured frequency responses. Listed ranges are source or assignment ranges, not confidence intervals; a dash means none is supplied. Volume fraction excludes background. Distance is the stored minimum to an electrode over sampled compartment voxels (at most 40,000), not skin depth or a guaranteed exact geometric minimum. Row order is preserved from the source (volume fraction divided by squared distance, with a 1 mm floor).

Display values are rounded to four significant figures. The unmodified full-precision table, assigned tissue classes and assignment notes are included as \texttt{table1-data.csv}. Label 52's corrupted display spelling is rendered as Skull Diploe. The source notes for labels 85 and 86 retain an obsolete \(10^{-15}\) S/m reference; their numeric assignments, used here and in the model, are \(10^{-6}\) S/m.

{\footnotesize\setlength{\tabcolsep}{3pt}
\def\LTcaptype{none}
\begin{longtable}{@{}>{\raggedright\arraybackslash}p{(\linewidth - 12\tabcolsep) * \real{0.045}}
>{\raggedright\arraybackslash}p{(\linewidth - 12\tabcolsep) * \real{0.29}}
>{\raggedright\arraybackslash}p{(\linewidth - 12\tabcolsep) * \real{0.105}}
>{\raggedright\arraybackslash}p{(\linewidth - 12\tabcolsep) * \real{0.2}}
>{\raggedright\arraybackslash}p{(\linewidth - 12\tabcolsep) * \real{0.065}}
>{\raggedright\arraybackslash}p{(\linewidth - 12\tabcolsep) * \real{0.145}}
>{\raggedright\arraybackslash}p{(\linewidth - 12\tabcolsep) * \real{0.15}}@{}}
\toprule
ID & Compartment & $\sigma$ (S/m) & Listed range (S/m) & Source & Volume fraction & Sampled distance (mm) \\
\midrule\endfirsthead
\multicolumn{7}{l}{\textit{Table 1 continued}}\\
\toprule
ID & Compartment & $\sigma$ (S/m) & Listed range (S/m) & Source & Volume fraction & Sampled distance (mm) \\
\midrule\endhead
\midrule\multicolumn{7}{r}{\textit{Continued on next page}}\\\endfoot
\bottomrule\endlastfoot
51 & Epidermis/Dermis & 0.465 & --- & S & 0.05962 & 0.7 \\
62 & Subcutaneous Adipose Tissue & 0.025 & --- & S & 0.09749 & 1.65 \\
43 & Adipose Tissue & 0.025 & --- & S & 0.1118 & 2.45 \\
63 & Muscle - Temporalis/Temporoparietalis & 0.355 & --- & S & 0.02868 & 2.55 \\
40 & Skull & 0.008 & --- & S & 0.04598 & 3.94 \\
38 & Muscle (General) & 0.355 & --- & S & 0.05825 & 5.19 \\
67 & Muscle - Splenius Capitis & 0.355 & --- & S & 0.008552 & 2.3 \\
68 & Muscle - Sternocleidomastoid & 0.355 & --- & S & 0.009119 & 2.55 \\
10 & Brain Gray Matter & 0.275 & --- & S & 0.1101 & 11.19 \\
54 & Skull Outer Table & 0.008 & --- & S & 0.02008 & 5.5 \\
12 & Brain White Matter & 0.126 & --- & S & 0.1169 & 14.06 \\
35 & Ear Auricular Cartilage (Pinna) & 0.17 & --- & S & 0.001851 & 1.87 \\
52 & Skull Diploe & 0.025 & --- & S & 0.02604 & 8.08 \\
32 & CSF General & 1.79 & --- & S & 0.04623 & 10.78 \\
36 & Mandible & 0.006302 & 0.001852--0.0091 & I & 0.01039 & 5.53 \\
89 & Parotid Gland & 0.5586 & 0.4356--0.743 & I & 0.01111 & 6.4 \\
53 & Skull Inner Table & 0.008 & --- & S & 0.0161 & 8.03 \\
60 & Muscle - Platysma & 0.355 & --- & S & 0.002112 & 3.05 \\
1 & Dura & 0.06 & 0.0006--0.06 & I & 0.02021 & 9.76 \\
66 & Muscle - Masseter & 0.355 & --- & S & 0.01241 & 8.27 \\
72 & Muscle - Depressor Anguli Oris & 0.355 & --- & S & 0.001387 & 4.39 \\
98 & Tendon - Temporalis Tendon & 0.3676 & 0.3676--0.3676 & I & 0.001796 & 5.24 \\
37 & Mucosa & 0.461 & 0.1--0.726 & I & 0.01077 & 13.51 \\
2 & Cerebellum Gray Matter & 0.5766 & 0.48--0.6465 & I* & 0.02004 & 19.16 \\
25 & Blood Veins & 0.7 & --- & S & 0.007009 & 11.89 \\
31 & Air Internal - Nasal/Pharynx & \(1\!\times\!10^{-6}\) & --- & S & 0.0103 & 15.1 \\
73 & Muscle - Depressor Labii & 0.355 & --- & S & 0.000703 & 4.35 \\
42 & Tongue & 0.355 & --- & S & 0.01537 & 25.5 \\
41 & Teeth & 0.006302 & 0.001852--0.0091 & I & 0.004493 & 14.68 \\
65 & Muscle - Lateral Pterygoid & 0.355 & --- & S & 0.004071 & 14.04 \\
9 & Cerebellum White Matter & 0.5766 & 0.48--0.6465 & I* & 0.01102 & 24.46 \\
84 & Muscle - Buccinator & 0.355 & --- & S & 0.00209 & 11.06 \\
88 & Submandibular Gland & 0.5586 & 0.4356--0.743 & I & 0.007198 & 21.79 \\
24 & Blood Arteries & 0.7 & --- & S & 0.003119 & 14.6 \\
97 & Air Internal - Oral Cavity & \(1\!\times\!10^{-6}\) & --- & S & 0.005035 & 19.7 \\
71 & Muscle - Mentalis & 0.355 & --- & S & 0.000264 & 5.14 \\
75 & Muscle - Orbicularis Oris & 0.355 & --- & S & 0.002573 & 17.57 \\
61 & Tendon - Galea Aponeurotica & 0.3676 & 0.3676--0.3676 & I & 0.009629 & 35.75 \\
81 & Muscle - Medial Pterygoid & 0.355 & --- & S & 0.004639 & 28.94 \\
85 & Ear Auditory Canal & \(1\!\times\!10^{-6}\) & \(1\!\times\!10^{-6}\)--\(1\!\times\!10^{-6}\) & S* & 0.000326 & 7.76 \\
83 & Muscles - Risorius & 0.355 & --- & S & 0.001301 & 15.97 \\
78 & Muscle - Zygomaticus Major & 0.355 & --- & S & 0.000877 & 14.11 \\
90 & Sublingual Gland & 0.5586 & 0.4356--0.743 & I & 0.001167 & 16.47 \\
44 & Vertebra - C1 (atlas) & 0.006302 & 0.001852--0.0091 & I & 0.002563 & 32.04 \\
80 & Muscle - Levator Scapulae & 0.355 & --- & S & 0.00312 & 37.36 \\
28 & Air Internal - Maxillary Sinus & \(1\!\times\!10^{-6}\) & --- & S & 0.002866 & 38.48 \\
6 & CSF Ventricles & 1.79 & --- & S & 0.002828 & 40.4 \\
108 & Cranial Nerve V3 - Mandibular Division & 0.348 & 0.06435--0.8524 & I & 0.000178 & 10.32 \\
45 & Vertebra - C2 (axis) & 0.006302 & 0.001852--0.0091 & I & 0.003609 & 46.69 \\
79 & Muscle - Orbicularis Oculi & 0.355 & --- & S & 0.002336 & 42.38 \\
57 & Eye Vitreous & 1.5 & --- & S & 0.003046 & 48.53 \\
70 & Muscle - Trapezius & 0.355 & --- & S & 0.003802 & 54.53 \\
8 & Putamen & 0.4191 & 0.06--0.83 & I* & 0.002291 & 43.71 \\
116 & Thalamus & 0.475 & 0.475--0.475 & I & 0.003147 & 52.57 \\
30 & Air Internal - Mastoid & \(1\!\times\!10^{-6}\) & --- & S & 0.000331 & 17.13 \\
46 & Vertebra - C3 & 0.006302 & 0.001852--0.0091 & I & 0.001957 & 42.31 \\
47 & Vertebra - C4 & 0.006302 & 0.001852--0.0091 & I & 0.001857 & 41.6 \\
14 & Brainstem Pons & 0.5584 & 0.369--0.7478 & I & 0.003179 & 55.19 \\
48 & Vertebra - C5 & 0.006302 & 0.001852--0.0091 & I & 0.001935 & 44.13 \\
39 & Nasal Septum (Cartilage) & 0.17 & --- & S & 0.002536 & 53.87 \\
86 & Ear Pharyngotympanic Tube & \(1\!\times\!10^{-6}\) & \(1\!\times\!10^{-6}\)--0.461 & S* & 0.000275 & 19.25 \\
87 & Hyoid Bone & 0.006302 & 0.001852--0.0091 & I* & 0.000285 & 19.68 \\
11 & Brainstem Midbrain & 0.35 & 0.35--0.35 & I & 0.001911 & 52.96 \\
5 & Hippocampus & 0.4191 & 0.06--0.83 & I & 0.001181 & 42.14 \\
69 & Muscle - Occipitiofrontalis - Occipital Belly & 0.355 & --- & S & 0.001319 & 48.94 \\
77 & Muscle - Levator Labii Superioris & 0.355 & --- & S & 0.000665 & 39.88 \\
7 & Caudate Nucleus & 0.4191 & 0.06--0.83 & I* & 0.001439 & 60.08 \\
13 & Spinal Cord & 0.611 & 0.4973--0.7815 & I & 0.001373 & 59.95 \\
26 & Air Internal - Ethmoidal Sinus & \(1\!\times\!10^{-6}\) & --- & S & 0.0013 & 58.89 \\
15 & Brainstem Medulla & 0.357 & 0.357--0.357 & I & 0.001147 & 55.96 \\
4 & Amygdala & 0.4191 & 0.06--0.83 & I* & 0.000689 & 45.28 \\
49 & Intervertebral Discs & 0.7393 & 0.6001--1.434 & I & 0.000625 & 43.94 \\
17 & Globus Pallidus & 0.4191 & 0.06--0.83 & I* & 0.000701 & 50.98 \\
56 & Eye Retina/Choroid/Sclera & 1.5 & --- & S & 0.000605 & 47.61 \\
93 & Muscle - Lateral Rectus & 0.355 & --- & S & 0.000321 & 43.46 \\
82 & Muscle - Zygomaticus Minor & 0.355 & --- & S & 0.000202 & 34.83 \\
74 & Muscle - Nasalis & 0.355 & --- & S & 0.000502 & 55.05 \\
64 & Muscle - Occipitiofrontalis - Frontal Belly & 0.355 & --- & S & 0.000978 & 79.26 \\
29 & Air Internal - Sphenoidal Sinus & \(1\!\times\!10^{-6}\) & --- & S & 0.000422 & 61.16 \\
92 & Muscle - Medial Rectus & 0.355 & --- & S & 0.000348 & 56.19 \\
27 & Air Internal - Frontal Sinus & \(1\!\times\!10^{-6}\) & --- & S & 0.000734 & 82.52 \\
91 & Muscle - Superior Rectus & 0.355 & --- & S & 0.000287 & 51.93 \\
94 & Muscle - Inferior Rectus & 0.355 & --- & S & 0.000257 & 51.05 \\
34 & Ear Semicircular Canals & 1.879 & 1.13--3.187 & I* & \(8.1\!\times\!10^{-5}\) & 29.26 \\
110 & Cranial Nerve VII - Facial & 0.348 & 0.06435--0.8524 & I & \(3.9\!\times\!10^{-5}\) & 24.13 \\
111 & Cranial Nerve VIII - Vestibulocochlear & 0.348 & 0.06435--0.8524 & I & \(8.6\!\times\!10^{-5}\) & 36.94 \\
103 & Cranial Nerve II - Optic & 0.348 & 0.06435--0.8524 & I & 0.000181 & 53.91 \\
112 & Cranial Nerve IX - Glossopharyngeal & 0.348 & 0.06435--0.8524 & I & \(7\!\times\!10^{-5}\) & 34.89 \\
113 & Cranial Nerve X - Vagus & 0.348 & 0.06435--0.8524 & I & \(7.2\!\times\!10^{-5}\) & 35.79 \\
16 & Nucleus Accumbens & 0.4191 & 0.06--0.83 & I* & 0.00022 & 63.6 \\
59 & Eye Aqueous & 1.5 & --- & S & 0.000194 & 62.68 \\
21 & Hypothalamus & 0.4191 & 0.06--0.83 & I & 0.000234 & 69.16 \\
95 & Muscle - Superior Oblique & 0.355 & --- & S & 0.000162 & 61.72 \\
96 & Muscle - Inferior Oblique & 0.355 & --- & S & 0.000106 & 51.13 \\
100 & Cerebral Peduncles & 0.348 & 0.06435--0.8524 & I* & 0.000105 & 55.28 \\
33 & Ear Cochlea & 1.879 & 1.13--3.187 & I* & \(4.5\!\times\!10^{-5}\) & 36.44 \\
106 & Cranial Nerve V - Trigeminal & 0.348 & 0.06435--0.8524 & I & \(9.6\!\times\!10^{-5}\) & 53.98 \\
22 & Commissura (Anterior) & 0.348 & 0.06435--0.8524 & I & \(8.4\!\times\!10^{-5}\) & 51.03 \\
107 & Cranial Nerve V2 - Maxillary Division & 0.348 & 0.06435--0.8524 & I & \(6\!\times\!10^{-5}\) & 46.36 \\
105 & Cranial Nerve IV - Trochlear & 0.348 & 0.06435--0.8524 & I & \(7.2\!\times\!10^{-5}\) & 51.8 \\
19 & Hypophysis or Pituitary Gland & 1.052 & 1.052--1.052 & I & 0.000106 & 64.71 \\
99 & Substantia Nigra & 0.4191 & 0.06--0.83 & I* & \(9.4\!\times\!10^{-5}\) & 63.44 \\
104 & Cranial Nerve III - Oculomotor & 0.348 & 0.06435--0.8524 & I & \(5.3\!\times\!10^{-5}\) & 50.25 \\
109 & Cranial Nerve VI - Abducens & 0.348 & 0.06435--0.8524 & I & \(4.4\!\times\!10^{-5}\) & 46.16 \\
101 & Optic Chiasm & 0.348 & 0.06435--0.8524 & I* & \(8.5\!\times\!10^{-5}\) & 64.16 \\
58 & Eye Cornea & 1.5 & --- & S & \(6.8\!\times\!10^{-5}\) & 63.49 \\
55 & Eye Lens & 1.5 & --- & S & \(6.6\!\times\!10^{-5}\) & 63.59 \\
76 & Muscles - Procerus & 0.355 & --- & S & \(9.9\!\times\!10^{-5}\) & 81.49 \\
18 & Optic Tract & 0.348 & 0.06435--0.8524 & I* & \(4.4\!\times\!10^{-5}\) & 61.95 \\
102 & Cranial Nerve I - Olfactory & 0.348 & 0.06435--0.8524 & I & \(3.2\!\times\!10^{-5}\) & 64.51 \\
20 & Mammillary Body & 0.4191 & 0.06--0.83 & I* & \(3.8\!\times\!10^{-5}\) & 71.02 \\
114 & Cranial Nerve XI - Accessory & 0.348 & 0.06435--0.8524 & I & \(1.4\!\times\!10^{-5}\) & 45.86 \\
3 & Pineal Body & 0.4811 & 0.4811--0.4811 & I & \(2.6\!\times\!10^{-5}\) & 73.79 \\
115 & Cranial Nerve XII - Hypoglossal & 0.348 & 0.06435--0.8524 & I & \(4\!\times\!10^{-6}\) & 52.69 \\
23 & Commissura (Posterior) & 0.348 & 0.06435--0.8524 & I & \(7\!\times\!10^{-6}\) & 75.06 \\
50 & Background & \(1\!\times\!10^{-6}\) & --- & S & 0 & --- \\
\end{longtable}
}

\textbf{Table 2, Tissue layer stack beneath each canonical site.} Millimetres per
MIDA tissue along the ray from each electrode through the full thickness of its
target. \emph{Target thickness traversed} is summed from \texttt{results/02\_layer\_profile.csv}
over the target label. \emph{Fat before target} is derived from
\texttt{results/02\_path\_composition.csv} as the adipose percentage of the
electrode-to-target path multiplied by that path length, i.e.~adipose
encountered \textbf{before} reaching the target, not over the whole ray.

{\def\LTcaptype{none} 
\begin{longtable}[]{@{}llll@{}}
\toprule\noalign{}
Site & Target & Target thickness traversed & Fat before target \\
\midrule\noalign{}
\endhead
\bottomrule\noalign{}
\endlastfoot
\texttt{submaxillary} & Mandible & 27.25 mm & 2.85 mm \\
\texttt{pre\_tragus} & Masseter & 17.25 mm & 5.68 mm \\
\texttt{midjaw} & Masseter & 16.50 mm & 9.05 mm \\
\texttt{submental\_lat} & Mandible & 11.00 mm & 1.50 mm \\
\texttt{buccal} & Buccinator & 8.00 mm & 8.62 mm \\
\texttt{submental\_mid} & Mandible & 7.25 mm & 7.04 mm \\
\texttt{hyoid} & Hyoid Bone & 6.25 mm & 14.01 mm \\
\texttt{above\_ear} & Temporalis & 6.00 mm & 2.00 mm \\
\texttt{mental} & Mentalis & 2.75 mm & 4.87 mm \\
\end{longtable}
}

\textbf{Table 3, Error budget.} The montage comparisons use ratios. A factor
that scales both lead fields equally cancels in their ratio; a factor that
varies between sites can affect it. Absolute-field effects are reported
separately.

{\def\LTcaptype{none} 
\begin{longtable}[]{@{}
  >{\raggedright\arraybackslash}p{(\linewidth - 12\tabcolsep) * \real{0.0400}}
  >{\raggedright\arraybackslash}p{(\linewidth - 12\tabcolsep) * \real{0.1400}}
  >{\raggedright\arraybackslash}p{(\linewidth - 12\tabcolsep) * \real{0.1900}}
  >{\raggedright\arraybackslash}p{(\linewidth - 12\tabcolsep) * \real{0.1100}}
  >{\raggedright\arraybackslash}p{(\linewidth - 12\tabcolsep) * \real{0.1700}}
  >{\raggedright\arraybackslash}p{(\linewidth - 12\tabcolsep) * \real{0.3500}}@{}}
\toprule\noalign{}
\begin{minipage}[b]{\linewidth}\raggedright
\#
\end{minipage} & \begin{minipage}[b]{\linewidth}\raggedright
Term
\end{minipage} & \begin{minipage}[b]{\linewidth}\raggedright
What sets it
\end{minipage} & \begin{minipage}[b]{\linewidth}\raggedright
Affects absolute
\end{minipage} & \begin{minipage}[b]{\linewidth}\raggedright
Affects ratios
\end{minipage} & \begin{minipage}[b]{\linewidth}\raggedright
Value
\end{minipage} \\
\midrule\noalign{}
\endhead
\bottomrule\noalign{}
\endlastfoot
1 & Discretisation & finite element size & yes & partly; directional, unquantified & not separable from term 6 at current precision, and not separable from term 10 either: a 4.9 \% linear refinement moved gaps 1.850 dB against 1.554 dB for a plain rebuild \\
2 & Interface proximity & source near a conductivity boundary & yes & yes; directional, unquantified & requires a geometry decoupling eccentricity from interface distance; not measured \\
3 & Inferior boundary & MIDA's cut face & yes & yes; jaw sites, not ear; directional, unquantified & unquantified; direction known, magnitude not bounded \\
4 & Muscle anisotropy & $\sigma$ tensor vs scalar & \textbf{yes; \textasciitilde5 dB in medial pterygoid, \textasciitilde4.5 dB in SCM} & \textbf{no; below the floor} & statistic A: largest change to any gap is \textbf{\textminus{}0.085 dB} (SCM); medial pterygoid \textminus{}0.010, temporalis +0.137, lateral pterygoid +0.036, all under the 0.27 dB floor. The absolute lead field IS affected; the reclassification is specific to ratios. Tensor on 2 of 10 compartments, the rest NOT APPLIED \\
5 & Fibre orientation & n unknown in MIDA & yes & yes & per-muscle min--max envelope \\
6 & Electrode meshing & contact area from incidental surface triangulation & yes & \textbf{yes; per-site} & \textbf{0.27 dB, 95 \% CI {[}0.17, 0.65{]}, n = 6} \\
7 & Single anatomy & MIDA is one subject & yes & unknown; directional, unquantified & not quantifiable from one head \\
8 & Delivered current & injected vs requested per solve & yes & \textbf{corrected for projected comparisons only} & 0.887--1.075 \texttimes{} requested across 22 solves, measured per solve by the tet-patch integral. Orientation-projected comparisons divide each site's field by its delivered current (\S{}2.4). The magnitude-only distance display and first cascade stage do not; their ratios retain this term. The 1.67 dB across-site spread is six times the row-6 floor and cannot be bounded by it \\
9 & Adipose conductivity & fat at 0.025 vs muscle 0.355 S/m & yes & \textbf{yes; and the SIGN differs by muscle} & Reported \textbf{per muscle only}; a population differential across sites has no clean definition under statistic A (the median change over muscles is +0.01 dB and conceals a sign that spans \textminus{}2.86 to +1.09). Statistic A, per muscle, sign varies: \textbf{\textminus{}1.121 dB} for temporalis (acts against the gap), \textbf{+0.411 dB} for sternocleidomastoid and \textbf{+0.323 dB} for lateral pterygoid (act with it), \textbf{\textminus{}2.863 to +1.093 dB} across the labial group. No single figure is admissible; the sign differs by muscle. Shares are not quoted here because they are dominated by the denominator; see \S{}4.3. \\
10 & Mesh realisation & which tetrahedra \texttt{meshmesh} produces from the same label volume & yes & \textbf{yes; measured, not bounded} & Rebuilding the mesh at the SAME nominal resolution on a second machine moves per-muscle gaps by up to \textbf{1.554 dB}. Reported as unquantified because this is ONE realisation pair, not a distribution; a bound needs several rebuilds. Not an estimator artefact: a volume-weighted mean moves 1.809 dB against the median's 1.850 dB on the same solved fields, so the movement is in the field. Wider than the envelopes of masseter, medial pterygoid, sternocleidomastoid and lateral pterygoid; the five labial verdicts, at 8.11 to 22.40 dB, are unaffected. \\
\end{longtable}
}

Row 6 is measured by rotating the electrode array and the source points together
on a fixed mesh, which preserves every source-to-electrode vector (verified to
2.8 \texttimes{} 10\textsuperscript{\textminus14} mm) so the exact answer is identical across draws and the whole
spread is realisation noise. The term splits into a common-mode part (SD
4.93 pp, 0.42 dB, cancels in a ratio) and an electrode-specific residual (SD
3.18 pp, 0.27 dB, does not). The reported statistic is the mean over 16
electrodes of the per-electrode SD across 6 draws, with a chi-square interval at
df = 5.

Row 7 is deliberately left unquantified. A single-subject model cannot estimate
its own between-subject variance.

\textbf{Table 4, Which montage sees which muscle, on two robustness axes.} Gap in dB
between the jaw and retroauricular montages, positive favouring the jaw, taken
as the median over 200 source orientations of the per-orientation gap (statistic
A) against the four pre-registered retroauricular sites. Electrode counts are
matched at 4 per montage. Because 5 jaw sites are admissible and the comparison
takes 4, every value is reported as the envelope over all 5 admissible subsets
and not for one chosen subset. Rows marked unstable change verdict between
subsets and are described in the text. \emph{Site-robust} asks whether a random draw
of four of the fourteen ear sites still excludes zero. \emph{Orientation agreement}
is the fraction of sampled orientations agreeing with the median verdict. The
temporalis verdict is specific to this basis and does not hold under the derived
fibre field; see \S{}4.8 and Figure 11.

{\def\LTcaptype{none} 
\begin{longtable}[]{@{}
  >{\raggedright\arraybackslash}p{(\linewidth - 10\tabcolsep) * \real{0.2000}}
  >{\raggedright\arraybackslash}p{(\linewidth - 10\tabcolsep) * \real{0.1700}}
  >{\raggedright\arraybackslash}p{(\linewidth - 10\tabcolsep) * \real{0.1200}}
  >{\raggedright\arraybackslash}p{(\linewidth - 10\tabcolsep) * \real{0.1600}}
  >{\raggedright\arraybackslash}p{(\linewidth - 10\tabcolsep) * \real{0.3500}}@{}}
\toprule\noalign{}
\begin{minipage}[b]{\linewidth}\raggedright
Muscle
\end{minipage} & \begin{minipage}[b]{\linewidth}\raggedright
Gap (dB), envelope over subsets
\end{minipage} & \begin{minipage}[b]{\linewidth}\raggedright
Site-robust
\end{minipage} & \begin{minipage}[b]{\linewidth}\raggedright
Orientation agreement
\end{minipage} & \begin{minipage}[b]{\linewidth}\raggedright
Verdict
\end{minipage} \\
\midrule\noalign{}
\endhead
\bottomrule\noalign{}
\endlastfoot
mentalis & +20.98 to +22.40 & yes, all 5 & 100.0 \% & \textbf{jaw, robust on both axes} \\
depressor anguli oris & +14.70 to +15.98 & yes, all 5 & 100.0 \% & \textbf{jaw, robust on both axes} \\
buccinator & +9.45 to +10.39 & yes, all 5 & 100.0 \% & \textbf{jaw, robust on both axes} \\
platysma & +9.90 to +10.27 & yes, all 5 & 100.0 \% & \textbf{jaw, robust on both axes} \\
orbicularis oris & +8.11 to +9.02 & yes, all 5 & 100.0 \% & \textbf{jaw, robust on both axes} \\
masseter & +1.20 to +2.22 & yes, all 5 & 56.0--68.5 \% & \textbf{jaw, site-robust but orientation-dependent} \\
medial pterygoid & +0.86 to +1.33 & yes, all 5 & 59.5--66.5 \% & \textbf{jaw, site-robust but orientation-dependent} \\
sternocleidomastoid & \textminus{}2.53 to \textminus{}0.97 & 1 of 5 & 60.5--65.0 \% & \textbf{unstable across subsets} \\
lateral pterygoid & \textminus{}3.61 to \textminus{}1.53 & 1 of 5 & 65.5--72.0 \% & \textbf{unstable across subsets} \\
temporalis & \textminus{}5.42 to \textminus{}2.57 & yes, all 5 & 92.0--100.0 \% & \textbf{ear, robust on both axes} \\
\end{longtable}
}

No jaw site is preferred by the design over any other, so rather than select one
4-site subset the comparison is run over all 5 admissible subsets and reported as
an envelope. 5 of 10 muscles favour the jaw montage on both robustness axes in
every subset, spanning 8.11 to 22.40 dB. One muscle, temporalis, favours the
retroauricular montage on both axes on this basis, in every subset. That verdict
does not survive a change of basis. Statistic A takes the median over a uniform
orientation sweep, and under the per-voxel fibre field derived from the label
volume the same comparison gives \textminus{}1.15 dB with an interval of {[}\textminus{}1.45, +5.46{]}
that spans zero (\S{}4.8, Figure 11). The derived lateral-pterygoid interval remains entirely on the ear side at the fixed published jaw basis (\S{}3.1). Table 4 retains the uniform-sweep results; it must not be read as a complete derived-field verdict table.

8 muscles return an identical verdict in all 5 subsets. Two do not. Lateral
pterygoid gives no resolvable preference across four subsets (\textminus{}1.53 to \textminus{}1.59 dB)
and \textminus{}3.61 dB in the fifth; sternocleidomastoid gives \textminus{}0.97 to \textminus{}1.44 dB across
four and \textminus{}2.53 dB in the fifth. In both cases the deviating subset is the one
dropping \texttt{midjaw}, and in both cases the deviating value is site-robust but
orientation-dependent, with agreement of 72.0 and 65.0 per cent against the
90 per cent required for a robust preference. Neither reaches a retroauricular
preference on both axes, and neither is among the 5 muscles carrying the jaw
advantage. \texttt{midjaw} has the largest perpendicular clearance to the cut plane of
any jaw site at 63.76 mm; the instability follows from removing the site
furthest from the truncation, not one near it.

\textbf{Temporalis is reported here under the uniform orientation sweep, for
comparability with the other nine muscles.} Under the fibre field derived from
the label volume (\S{}2.5.1), its median is \textminus{}1.15 dB and its interval
over all 1001 four-site ear subsets is {[}\textminus{}1.45, +5.46{]} dB. This
interval crosses zero and governs the main-analysis conclusion (\S{}3.1,
\S{}4.1). The derivation replaces the uniform-orientation assumption with a
segmented-anatomy construction; it is not assumption-free. The table retains
the uniform-sweep treatment so that the difference between treatments can be
examined.

\textbf{Table 5, Strongest single site for each articulator, and how well resolved it
is.} For each of the ten segmented articulators, the electrode with the largest
projected lead field under the isotropic model, its runner-up, and the margin
between them. Margins marked * are at or below 0.65 dB, the upper bound of the
95 \% CI on the electrode-meshing floor (\S{}2.7, row 6), so for those muscles the
identity of the strongest site is not resolved by this model. For every
articulator, the strongest site and its runner-up belong to the same montage.
Uncertainty between those two sites therefore leaves their montage assignment
unchanged.
These are single electrodes under an unmatched argmax, so the table records
where coupling is strongest and is not a montage recommendation; see \S{}4.4.
\texttt{earlobe\_ipsi} is solved but belongs to no montage and is excluded.

{\def\LTcaptype{none} 
\begin{longtable}[]{@{}
  >{\raggedright\arraybackslash}p{(\linewidth - 12\tabcolsep) * \real{0.2000}}
  >{\raggedright\arraybackslash}p{(\linewidth - 12\tabcolsep) * \real{0.1800}}
  >{\raggedright\arraybackslash}p{(\linewidth - 12\tabcolsep) * \real{0.1000}}
  >{\raggedright\arraybackslash}p{(\linewidth - 12\tabcolsep) * \real{0.1800}}
  >{\raggedright\arraybackslash}p{(\linewidth - 12\tabcolsep) * \real{0.1300}}
  >{\raggedright\arraybackslash}p{(\linewidth - 12\tabcolsep) * \real{0.2100}}@{}}
\toprule\noalign{}
\begin{minipage}[b]{\linewidth}\raggedright
Articulator
\end{minipage} & \begin{minipage}[b]{\linewidth}\raggedright
Best single site
\end{minipage} & \begin{minipage}[b]{\linewidth}\raggedright
Montage
\end{minipage} & \begin{minipage}[b]{\linewidth}\raggedright
Runner-up
\end{minipage} & \begin{minipage}[b]{\linewidth}\raggedright
Margin (dB)
\end{minipage} & \begin{minipage}[b]{\linewidth}\raggedright
Runner-up crosses montage
\end{minipage} \\
\midrule\noalign{}
\endhead
\bottomrule\noalign{}
\endlastfoot
masseter & buccal & jaw & mental & 0.86 & no \\
temporalis & cg01 & ear & cg02 & 0.71 & no \\
medial pterygoid & hyoid & jaw & submaxillary & 0.88 & no \\
lateral pterygoid & pre-tragus & ear & cg10 & 0.85 & no \\
orbicularis oris & mental & jaw & buccal & 1.02 & no \\
buccinator & mental & jaw & submental (mid) & 0.88 & no \\
mentalis & mental & jaw & submental (mid) & 0.27 * & no \\
depressor anguli oris & submental (mid) & jaw & mental & 0.15 * & no \\
platysma & hyoid & jaw & submental (mid) & 0.37 * & no \\
sternocleidomastoid & post-lobule & ear & cg08 & 0.03 * & no \\
\end{longtable}
}

\begin{center}\rule{0.5\linewidth}{0.5pt}\end{center}

\noindent\includegraphics[width=\linewidth]{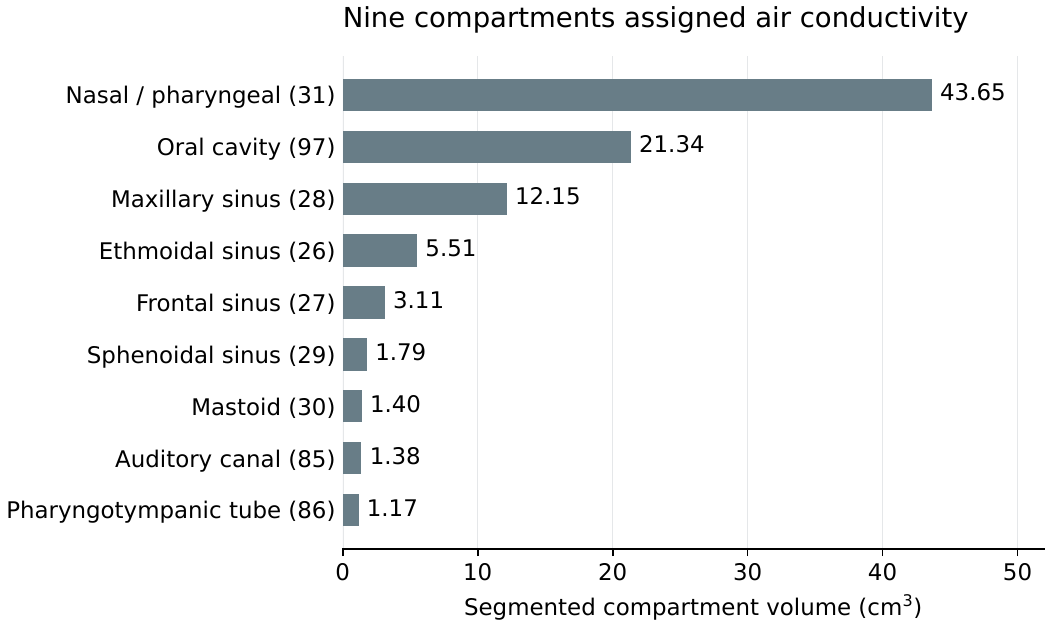}\par

\textbf{Supplementary Figure S1.} Segmented volumes of the nine compartments assigned air conductivity, excluding background (label 50). Values come from \texttt{results/01\_label\_inventory.csv}; label numbers are shown in parentheses. These are compartment volumes, not measurements of air content during articulation. The pharyngotympanic tube's air assignment is a modelling choice (Table 1). The earlier caption promised skin depth, but no corresponding depth output is supplied; this figure reports volume only, without estimating depth.

\begin{center}\rule{0.5\linewidth}{0.5pt}\end{center}

\FloatBarrier
\subsection{Data and code availability}\label{data-and-code-availability}

All analysis code, electrode definitions, conductivity assignments and result
tables are available at \texttt{github.com/Somach-Systems-Inc/ear-emg-forward-model}.
MIDA's licence prohibits redistributing the model. It must be obtained from
the IT'IS Foundation, as described in \texttt{REPRODUCTION.md}. Commits \texttt{fa583f6} (2026-08-02) and the lead-field commit (2026-08-03) are
cited in \S{}2.8 as the pre-registration record.

\begin{center}\rule{0.5\linewidth}{0.5pt}\end{center}

\FloatBarrier
\subsection{References}\label{references}

\begin{enumerate}
\def\labelenumi{\arabic{enumi}.}
\tightlist
\item
  An, et al.~(2025). ID.EARS: One-Ear EEG Device with Biosignal Noise for Real-Time Gesture Recognition and Various Interactions. \emph{CHI '25}, 1--18. doi:10.1145/3706598.3714185
\item
  Avramidou, et al.~(2024). From Ear-EEG to Ear-ExG: The Jaw Artifact is a Keeper. \emph{DSAI '24}.
\item
  Debener, S., et al.~(2015). Unobtrusive ambulatory EEG using a smartphone and flexible printed electrodes around the ear. \emph{Sci Rep} 5:16743. doi:10.1038/srep16743
\item
  De Luca, C. J., et al.~(2012). Inter-electrode spacing of surface EMG sensors. \emph{J Biomech} 45(3):555--561. doi:10.1016/j.jbiomech.2011.11.010
\item
  Gaddy, D., \& Klein, D. (2020). Digital Voicing of Silent Speech. \emph{EMNLP}, 5521--5530. doi:10.18653/v1/2020.emnlp-main.445
\item
  Goncharova, I. I., et al.~(2003). EMG contamination of EEG: spectral and topographical characteristics. \emph{Clin Neurophysiol}.
\item
  Harmening, N., Klug, M., Gramann, K., \& Miklody, D. (2022). HArtMuT, modeling eye and muscle contributors in neuroelectric imaging. \emph{J Neural Eng} 19(6):066041. doi:10.1088/1741-2552/aca8ce
\item
  Iacono, M. I., et al.~(2015). MIDA: A Multimodal Imaging-Based Detailed Anatomical Model of the Human Head and Neck. \emph{PLOS ONE}. doi:10.1371/journal.pone.0124126
\item
  Kappel, S. L., Makeig, S., \& Kidmose, P. (2019). Ear-EEG Forward Models: Improved Head-Models for Ear-EEG. \emph{Front Neurosci} 13:943. doi:10.3389/fnins.2019.00943
\item
  Kapur, A., Kapur, S., \& Maes, P. (2018). AlterEgo: A Personalized Wearable Silent Speech Interface. \emph{IUI '18}, 43--53. doi:10.1145/3172944.3172977
\item
  Kuiken, T. A., Lowery, M. M., \& Stoykov, N. S. (2003). The effect of subcutaneous fat on myoelectric signal amplitude and cross-talk. \emph{Prosthet Orthot Int} 27(1):48--54. doi:10.3109/03093640309167976
\item
  Maksymenko, K., Deslauriers-Gauthier, S., \& Farina, D. (2021). Ultra fast and highly realistic numerical modelling of surface EMG. \emph{bioRxiv}.
\item
  Meiser, A., Knoll, J., \& Bleichner, M. G. (2024). High-density ear-EEG for understanding ear-centered EEG. \emph{J Neural Eng} 21(1):016001. doi:10.1088/1741-2552/ad1783
\item
  Mesin, L. (2020). Crosstalk in surface electromyogram: literature review and some insights. \emph{Phys Eng Sci Med} 43(2):481--492. doi:10.1007/s13246-020-00868-1
\item
  Sato, W., \& Kochiyama, T. (2023). Crosstalk in Facial EMG and Its Reduction Using ICA. \emph{Sensors} 23:2720.
\item
  Saturnino, G. B., Madsen, K. H., \& Thielscher, A. (2019). Electric field simulations for transcranial brain stimulation using FEM: an efficient implementation and error analysis. \emph{J Neural Eng} 16(6):066032. doi:10.1088/1741-2552/ab41ba
\item
  Thielscher, A., Antunes, A., \& Saturnino, G. B. (2015). Field modeling for transcranial magnetic stimulation: a useful tool to understand the physiological effects of TMS? \emph{EMBC 2015}, 222--225. doi:10.1109/EMBC.2015.7318340
\item
  Wand, M., \& Schultz, T. (2011). Session-Independent EMG-Based Speech Recognition. \emph{BIOSIGNALS 2011}, 295--300. doi:10.5220/0003169702950300
\item
  Yao, D., et al.~(2019). Which Reference Should We Use for EEG and ERP practice? \emph{Brain Topogr}.
\item
  Yarici, M., Thornton, M., \& Mandic, D. P. (2023). Ear-EEG sensitivity modeling for neural sources and ocular artifacts. \emph{Front Neurosci} 16:997377. doi:10.3389/fnins.2022.997377
\end{enumerate}

\end{document}